\documentclass[useAMS,usenatbib]{mn2e}
\usepackage{amssymb,amsmath,epsfig,times, natbib,subfig,color}

\title[Perseus fluctuations]{What fraction of the density fluctuations in the Perseus cluster core is due to gas sloshing rather than AGN feedback?}
\author[S. A. Walker et al] {\parbox[]{6.5in}{{
      S. A. Walker,$^1$\thanks{Email:stephen.a.walker@nasa.gov} J. S. Sanders$^2$, A. C. Fabian$^3$ 
    }\\
     \footnotesize
     $^1$Astrophysics Science Division, X-ray Astrophysics Laboratory, Code 662, NASA Goddard Space Flight Center, Greenbelt, MD 20771, USA \\
   $^2$Max-Planck-Institute fur extraterrestrische Physik, 85748 Garching, \\
        $^3$Institute of Astronomy, Madingley Road, Cambridge CB3 0HA \\
    }}

\date{}

\begin{document}

\maketitle

\begin{abstract}
Deep Chandra observations of the core of the Perseus cluster show a plethora of complex structure. It has been found that when the observed density fluctuations in the intracluster medium are 
converted into constraints on AGN induced
turbulence, the resulting turbulent heating rates are sufficient to balance cooling locally throughout the central 220kpc. However while the signatures of AGN feedback (inflated bubbles) dominate
the central 60kpc in X-ray images, beyond this radius the intracluster medium is increasingly shaped by the effects of gas sloshing, which can also produce subtle variations in X-ray surface brightness.
We use mock Chandra observations of gas sloshing simulations to investigate what fraction of the observed density fluctuations in the core of the Perseus galaxy cluster may originate from sloshing rather than 
AGN induced feedback. Outside 60kpc, we find that the observed level of the density fluctuations is broadly consistent with being produced by sloshing alone. If this is the case, AGN-generated turbulence is likely to be insufficient in combating cooling outside 60kpc.

\end{abstract}


\section{Introduction}
Left unopposed, the cooling of the intracluster medium in the centres of cool-core galaxy clusters such as the Perseus cluster should lead to the formation 
of cooling flows of $\sim$100s M$_{\odot}$ yr$^{-1}$. Over the last two decades, the unrivalled spatial resolution of Chandra's imaging detectors, combined with the 
high spectral resolution of XMM-Newton's RGS, have shown that feedback from the central AGN in clusters counteracts this cooling, preventing the formation
of strong cooling flows (\citealt{Peterson2006}, \citealt{McNamara2007}, \citealt{Fabian2012}, \citealt{McNamara2012}).

Understanding the way AGN feedback in galaxy clusters is dissipated into the intracluster medium (ICM) is one of the most pressing issues in cluster astrophysics. How can the energy from the jets of central 
AGN be so uniformly spread throughout the hundreds of kiloparsecs spanned by the cool core, in a way that perfectly balances the rate of gas cooling?

As the brightest galaxy cluster in the X-ray sky, the Perseus cluster has been at the forefront in the advancement of our understanding of cluster astrophysics
and AGN feedback. It was in Perseus that AGN inflated cavities were first identified, using ROSAT (\citealt{Boehringer1993}). Deep Chandra observations have revealed
a plethora of structure, including multiple generations of cavity pairs, and ripples resembling sound waves (\citealt{Fabian2003,Fabian2006}, \citealt{Sanders2007}).

Recent observations with the high-spectral resolution microcalorimeter on Hitomi of the central regions of the Perseus cluster have found a remarkably
 low velocity dispersion of 164$\pm$10 km s$^{-1}$ (\citealt{Hitomi2016}). If this 
dispersion is interpreted as turbulence, it leads to a very low turbulent to thermal pressure ratio of only 4 percent. To support the ICM from cooling,
 this turbulence needs to be regenerated on a timescale of just 
4 percent of the cooling time, during which it can only travel less than 13kpc, far smaller than the $\approx$200kpc size of Perseus's cool core. For turbulence
 to be the mechanism dissipating heat into the ICM, it would
need to be generated `in situ' throughout the entire core. Attempts to simulate the generation of turbulence resulting from jets being fired
into the intracluster medium have found that only a very low level of turbulence (less that 1 percent of the thermal energy) develops (\citealt{Reynolds2015}, 
\citealt{Yang2016}, \citealt{Bourne2017}).

Alternatively, \citet{Fabian2017} have found that sound waves, which travel an order of magnitude faster
 and can cross the core in a cooling time, are also consistent with the Hitomi and
Chandra results. \citet{Bambic2018} have also found that the upper limits on the 
turbulent velocities obtained using XMM-Newton's RGS 
for a sample of 4 clusters are too low to allow energy to propagate radially to 
their cooling radii within the required cooling time.

Constraints on the level of turbulence have also been made using the power spectrum of density fluctuations seen in the ICM with Chandra (\citealt{ChurazovComa}, \citealt{SandersAWM7}).
\citet{Zhuravleva2014L} used cosmological simulations of relaxed galaxy clusters 
to investigate the relation between the observed rms density fluctuations (which can be readily observed in 
X-ray images), and the velocity fluctuations in these simulations. They found that the rms of the density fluctuations and the 
velocity fluctuations are linearly related across a wide range of scales, all the way down to 
the small scales of the turbulent regime. Using these cosmological simulations, \citet{Zhuravleva2014L}
found the proportionality coefficient between the two to be
consistent with unity (1$\pm$0.3).

\citet{Zhuravleva2014,Zhuravleva2015} have used these methods for the core of Perseus cluster.
If these density fluctuations are entirely due to AGN induced turbulence, \citet{Zhuravleva2014} found that the rate of dissipation
 of heat at each radius studied can match the cooling rate locally throughout the central 220kpc of the core of Perseus.  \citet{Zhuravleva2014} also found a 
similar balance for the core of the Virgo cluster, while \citet{Walker2015c} found the same in the Centaurus cluster. The velocities found by \citet{Zhuravleva2014} using this method in the region of the Perseus cluster studied
by Hitomi (30-60 kpc from the cluster's nucleus) on the spatial scales of the largest AGN inflated bubbles
in the field (20-30kpc) are consistent with the observed Hitomi value of 164$\pm$10 km s$^{-1}$.

However, while the signatures of AGN feedback dominate the central 60kpc of the Perseus cluster (with AGN inflated bubbles on either side
 of the core), beyond 60kpc the X-ray surface brightness distribution becomes dominated
by the characteristic spiral pattern brought about through gas sloshing (\citealt{Walker2017}, \citealt{Walker2018}). Such gas sloshing
 is also able to produce subtle fluctuations in X-ray emission 
(\citealt{ZuHone2011,ZuHone2018}, \citealt{Walker2017}), which need to be taken into account when converting the observed power spectrum of gas density
 fluctuations into constraints on the level of AGN induced turbulence. As the magnetic field wraps around a cold front, it can produce layers of higher and lower magnetic pressure, which reduce and increase the
observed X-ray surface brightness respectively (\citealt{Werner2016}). 

As described in \citet{Zhuravleva2014}, the current constraints on turbulent velocities resulting from AGN feedback using the density fluctuation method are upper limits, since other physical processes such as sloshing contribute. Here we investigate the power spectrum of X-ray surface brightness fluctuations seen in gas sloshing simulations resembling
 the sloshing seen in Perseus, to determine what fraction of the observed fluctuations can be explained by
sloshing. By extension this therefore allows us to understand how much of the density fluctuations in the Perseus cluster core 
are due to AGN feedback alone, and how far out any AGN driven turbulence can extend.

\begin{figure*}
  \begin{center}
    \leavevmode
\epsfig{file=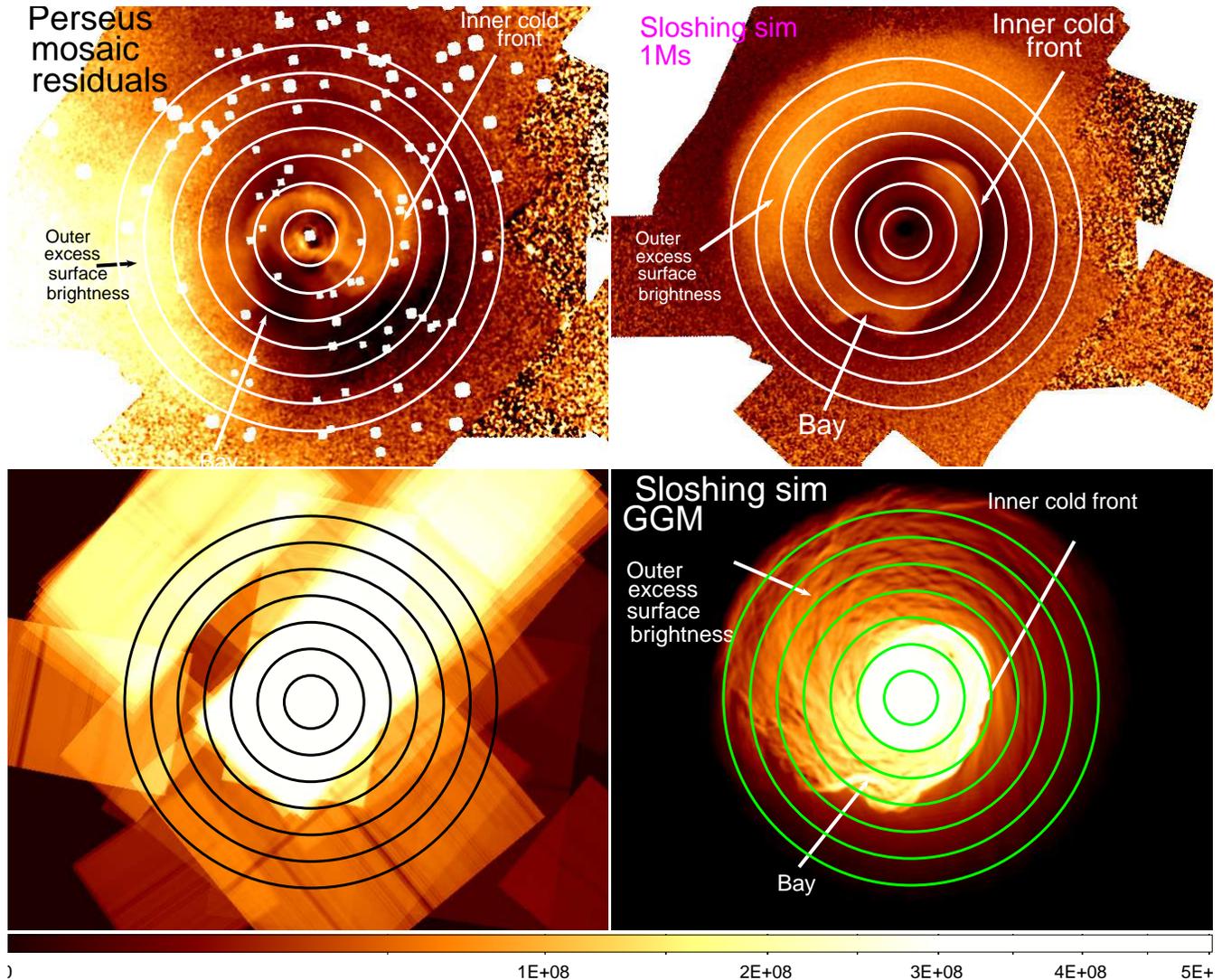, width=\linewidth}
      \caption{\emph{Top left}: azimuthal residuals of Perseus cluster with the 7 annuli used in \citet{Zhuravleva2014} overplotted.
\emph{Top right}:Sloshing simulation rotated to roughly match the Perseus inner spiral, with the same sized annuli overplotted and exposure 
map matching that of the actual Perseus mosaic. \emph{Bottom left}: Exposure map of Perseus mosaic used for making simulated images. The colourbar is for the exposure map and has units of cm$^2$ s$^{-1}$.
 \emph{Bottom right}:GGM filtered version of the input sloshing simulation with annuli overplotted. Annuli are in increments of 1.5arcmin ($\approx$30kpc) from 0-220kpc.}
      \label{Images}
  \end{center}
\end{figure*}

\section{Analysis: simple beta model removal method}
\label{sec:analysis_beta}
Here we perform the same X-ray density fluctuation analysis as \citet{Zhuravleva2014}, firstly on the real Perseus data, and then on mock Chandra observations of a sloshing simulation that provides
the best fit to the sloshing spiral seen in Perseus (\citealt{Walker2017}).

\subsection{Perseus data analysis}
\label{sec:analysis_beta:Perseus}
Firstly we repeat the analysis of \citet{Zhuravleva2014}, dividing the Perseus cluster into 7 annuli (Fig. \ref{Images}, top left), each with width 1.5 arcmins, 
which is $\approx$30 kpc. We used an exposure corrected mosaiced image in the same band (0.5-3.5keV). The Chandra observations used are listed here by obs-id: 3209, 4289, 4946, 4947, 4948, 4949, 4950, 4951, 4952, 4953,
6139, 6145, 6146, 11713, 11714, 11715, 11716, 12025, 12033,
12036, 12037. 
%for the assumed redshift of 0.01756. 
Point sources were excised from the image. The image was divided by the same $\beta$ model as in \citet{Zhuravleva2014}, 
to find the surface brightness residuals. We then use the delta variance technique (\citealt{Deltavariance}) to find the 2D power spectrum, P$_{2D}$(k), of surface
brightness fluctuations in each annuli as a function of the wavenumber, $k=1/l$. Following \citet{ChurazovComa}, this 2D power spectrum is 
converted into a 3D power spectrum, P$_{3D}$(k), using,

\begin{figure*}
  \begin{center}
 \hbox{
\epsfig{file=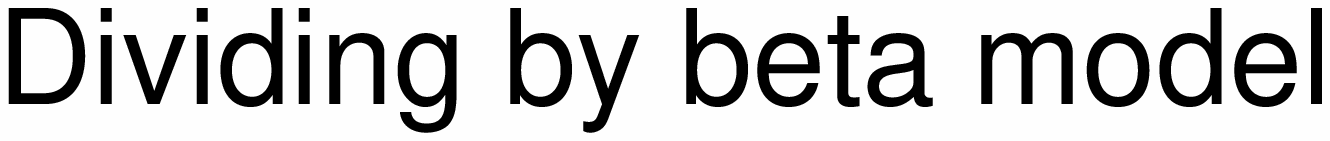, width=\linewidth}

}

 \hbox{
\epsfig{file=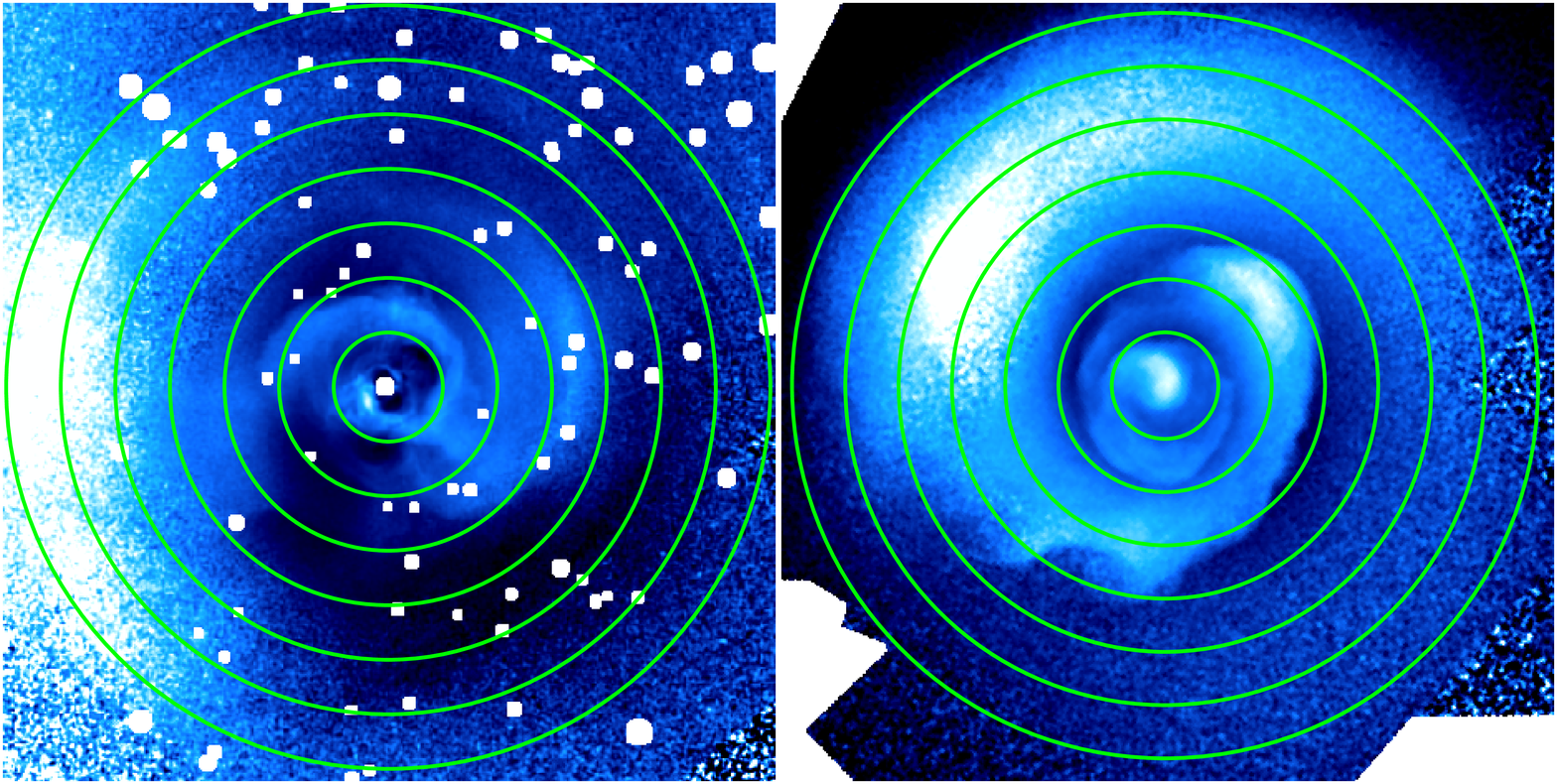, width=0.45\linewidth}
\hspace{1.0cm}
\epsfig{file=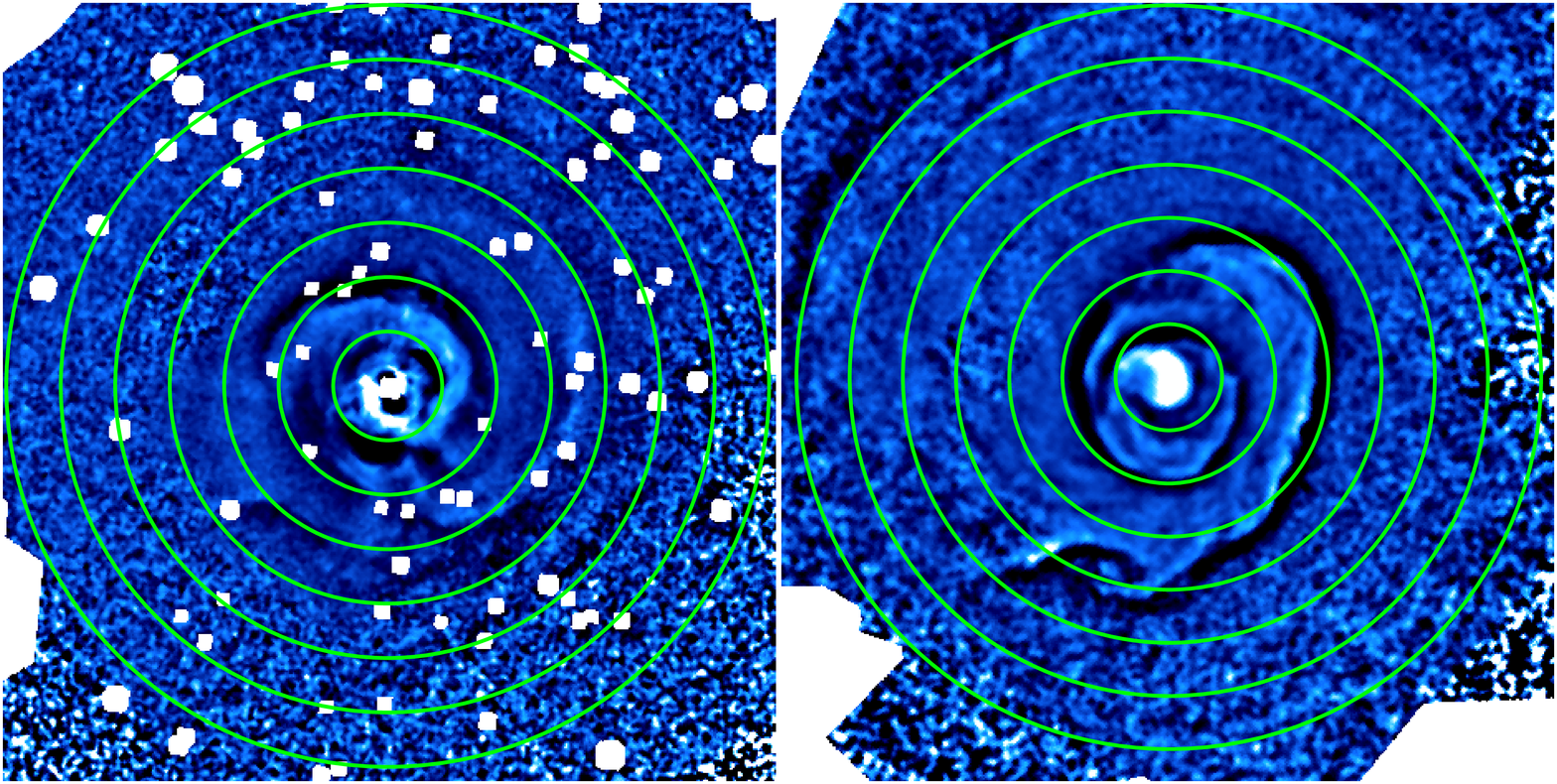, width=0.45\linewidth}

}

\vspace{-0.5cm}

\hbox{
\epsfig{file=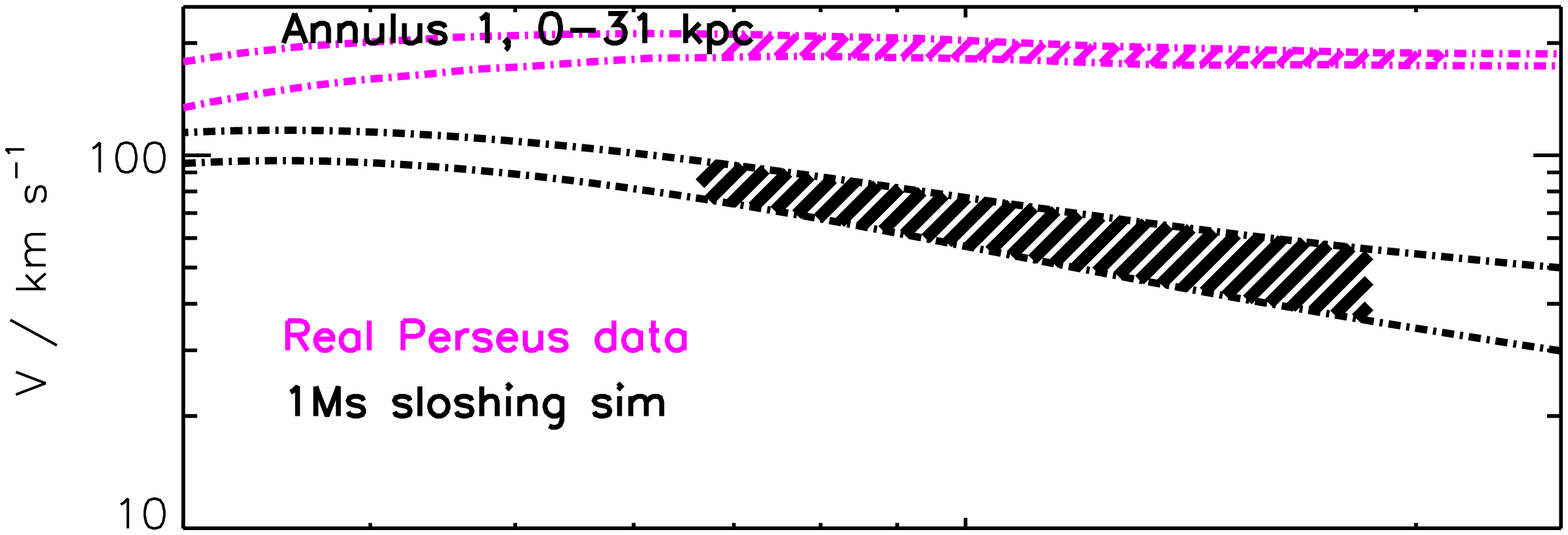, width=0.46\linewidth}
    \epsfig{file=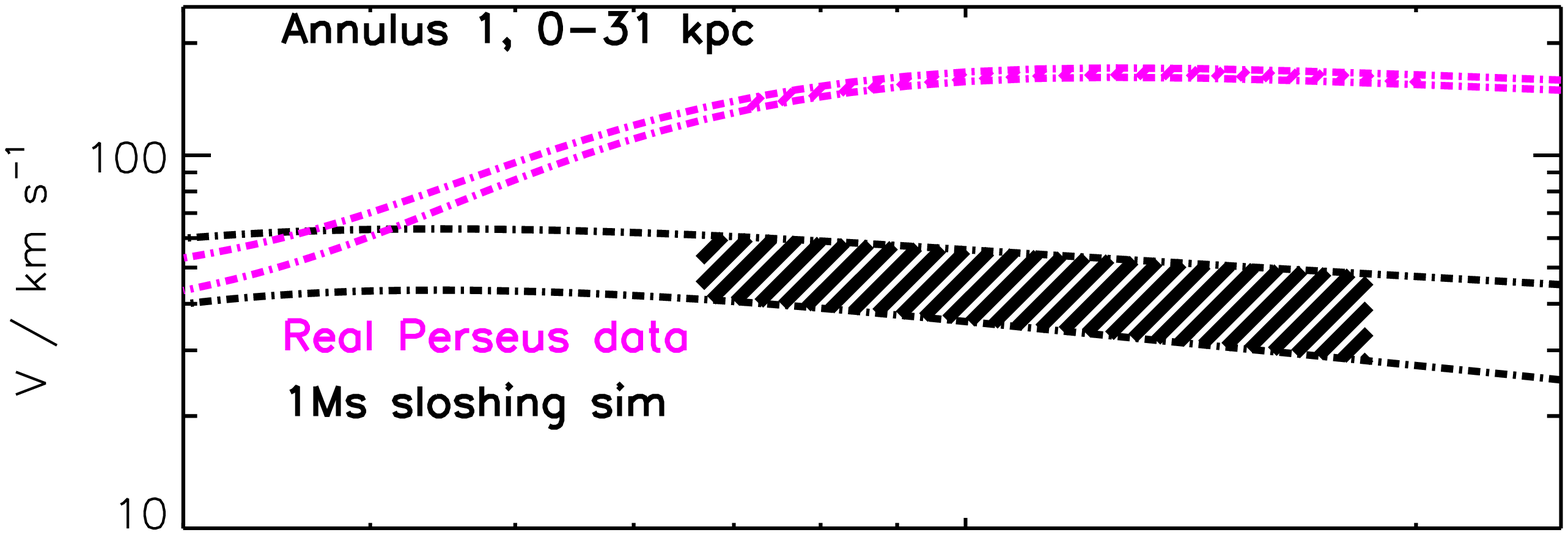, width=0.46\linewidth}
}  
\vspace{-1.75cm}

\hbox{
\epsfig{file=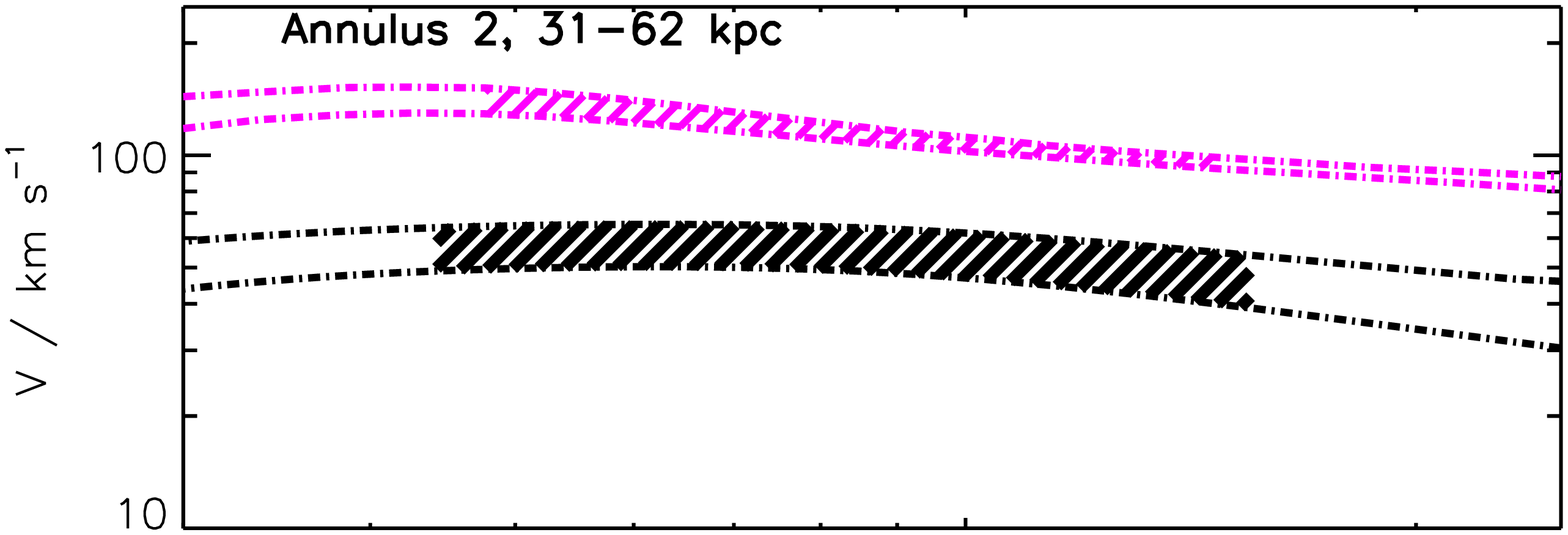, width=0.46\linewidth}
    \epsfig{file=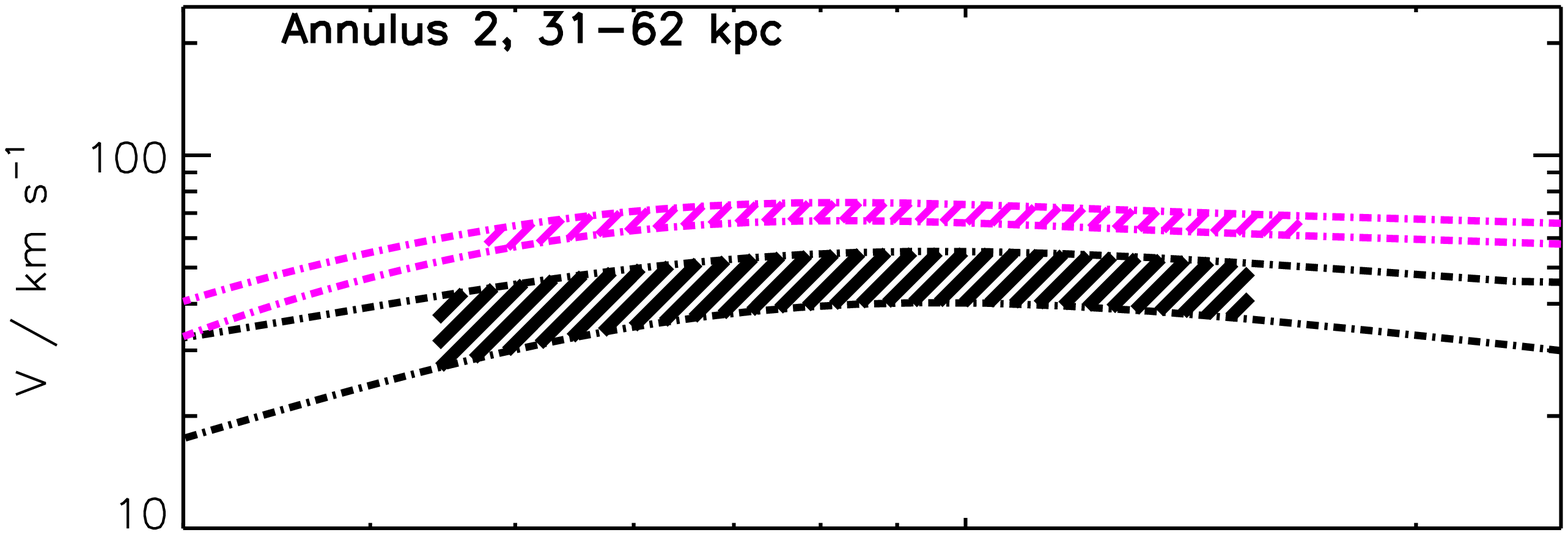, width=0.46\linewidth}
}  
\vspace{-1.75cm}

\hbox{
\epsfig{file=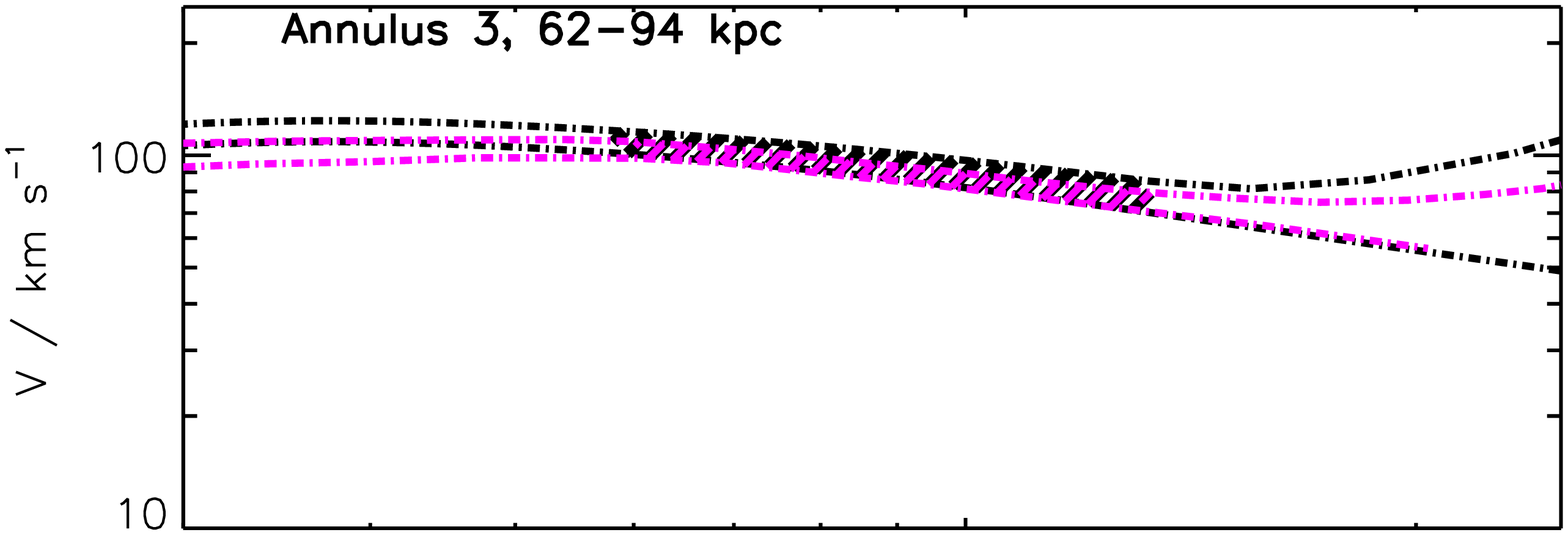, width=0.46\linewidth}
    \epsfig{file=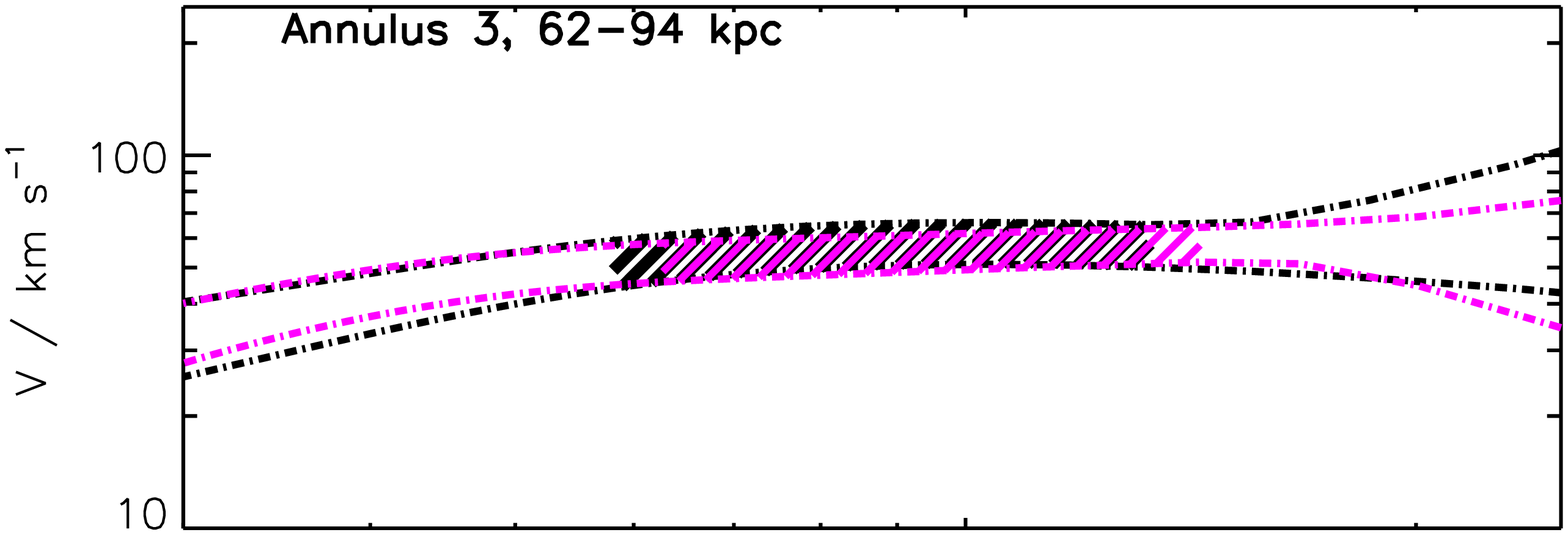, width=0.46\linewidth}
}  
\vspace{-1.75cm}

\hbox{
\epsfig{file=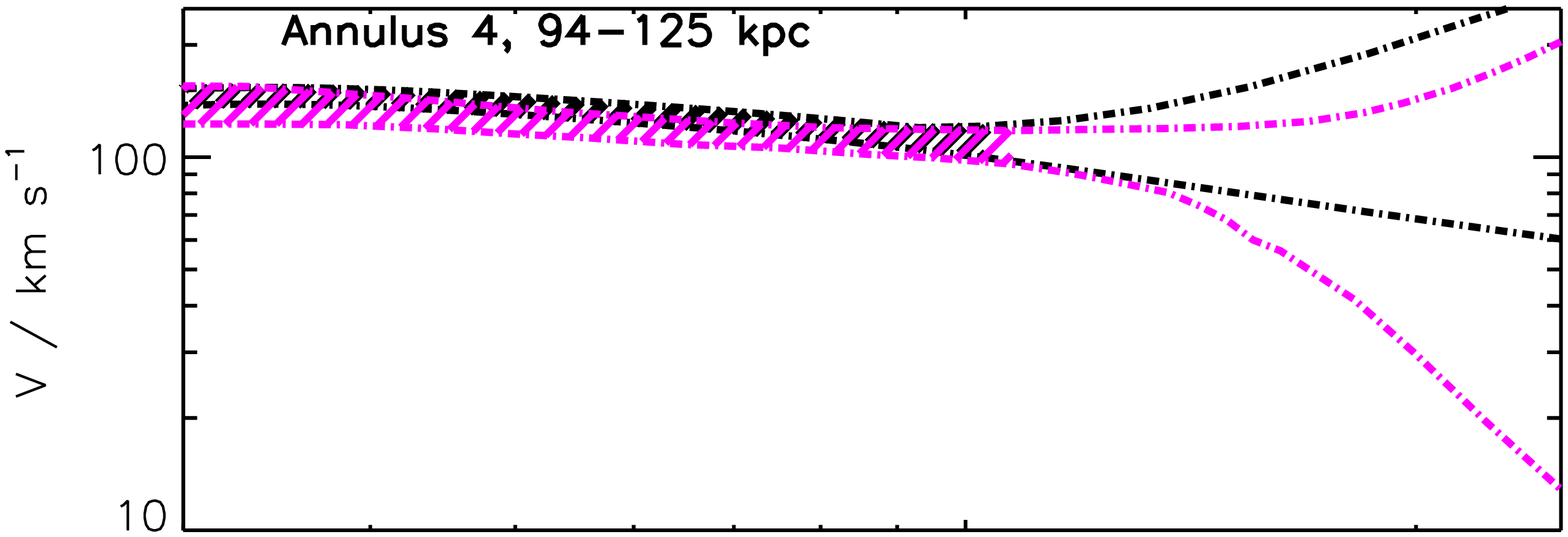, width=0.46\linewidth}
    \epsfig{file=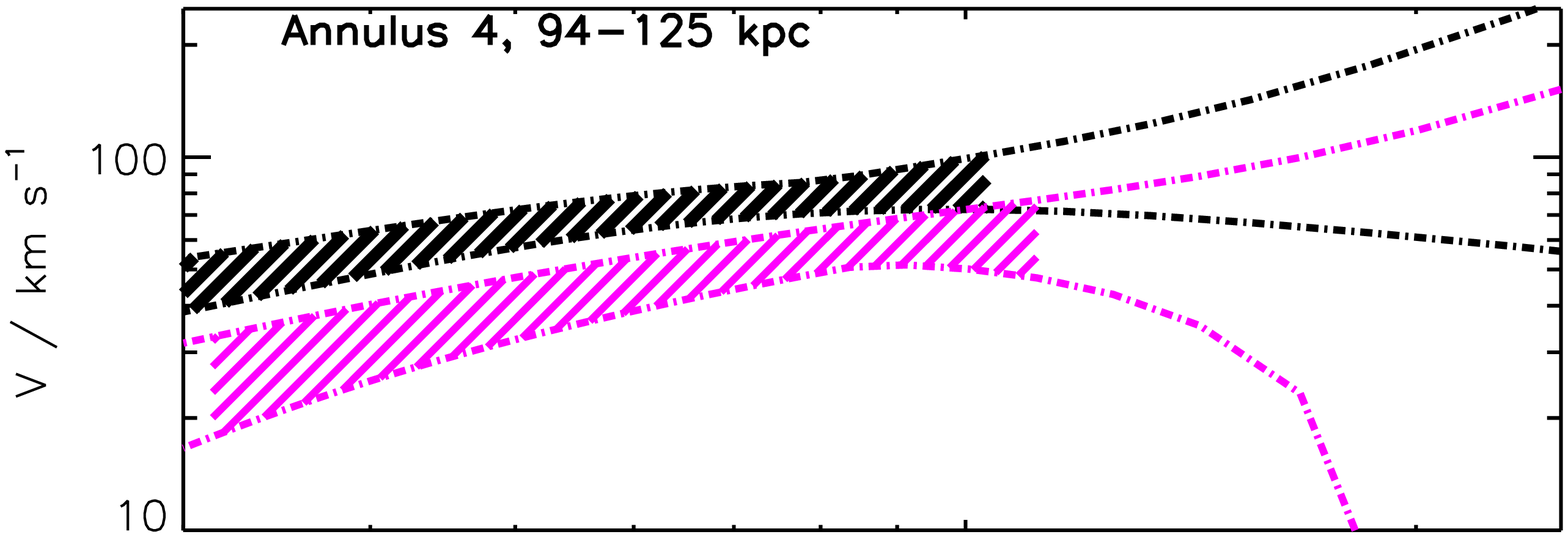, width=0.46\linewidth}
}  

\vspace{-1.75cm}

\hbox{
\epsfig{file=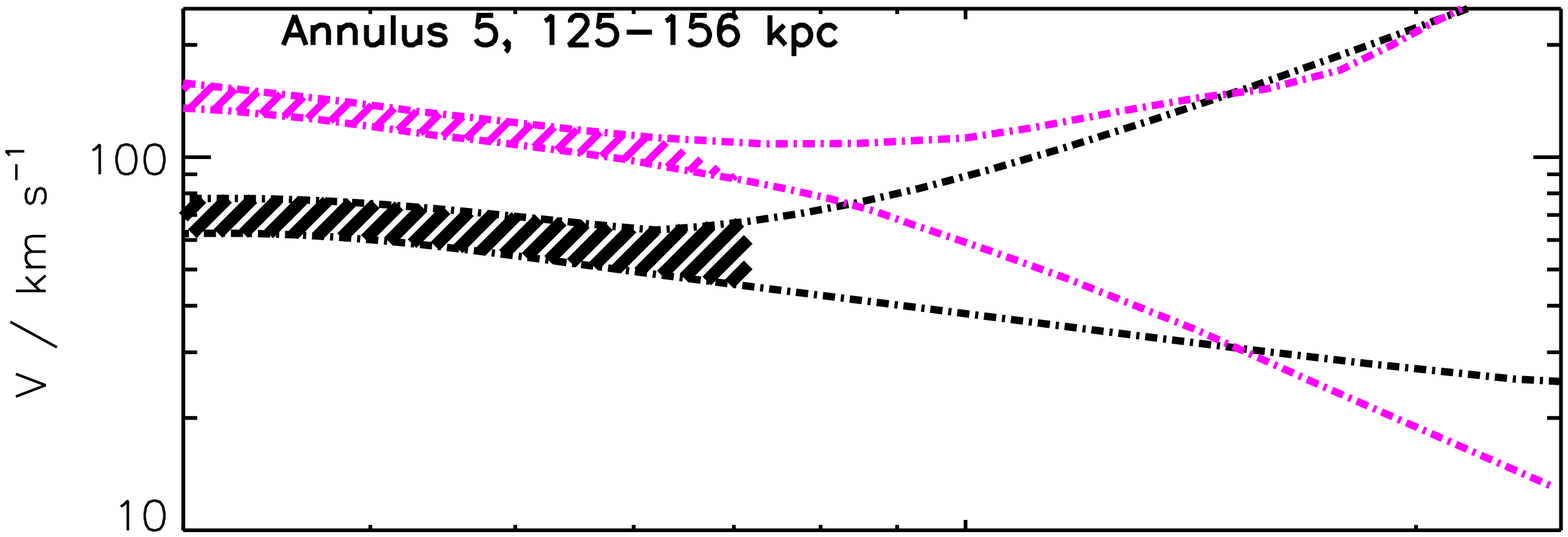, width=0.46\linewidth}
    \epsfig{file=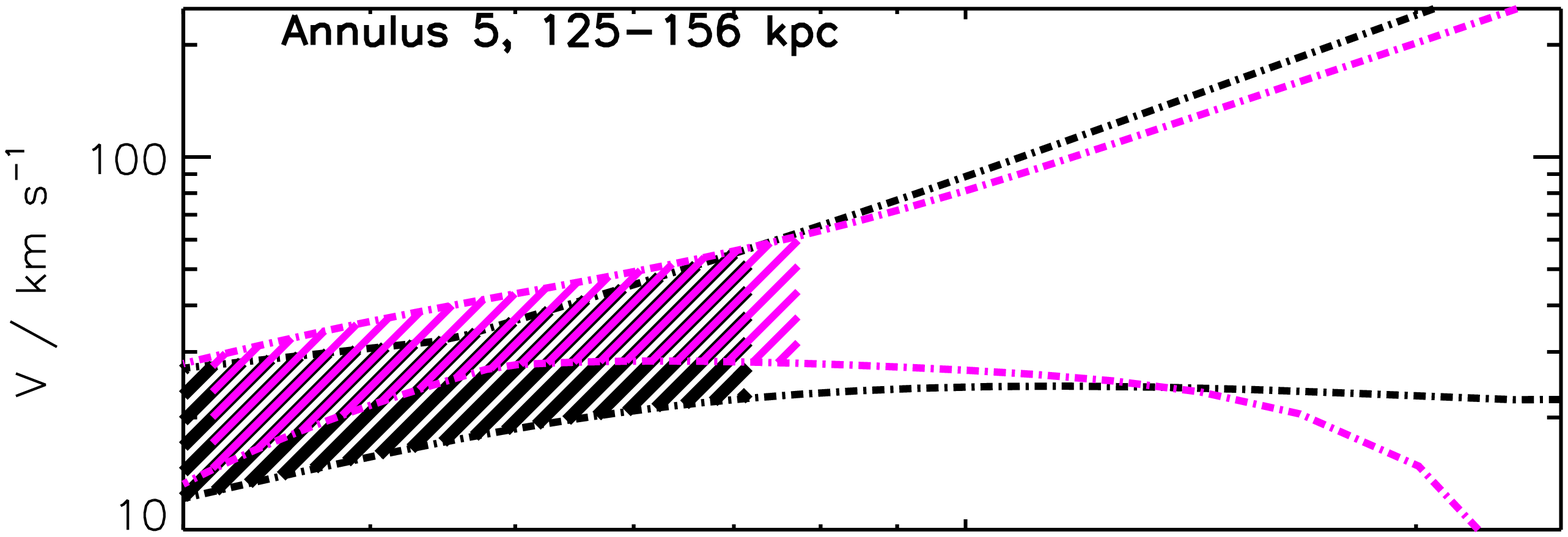, width=0.46\linewidth}
}
\vspace{-1.75cm}

\hbox{
\epsfig{file=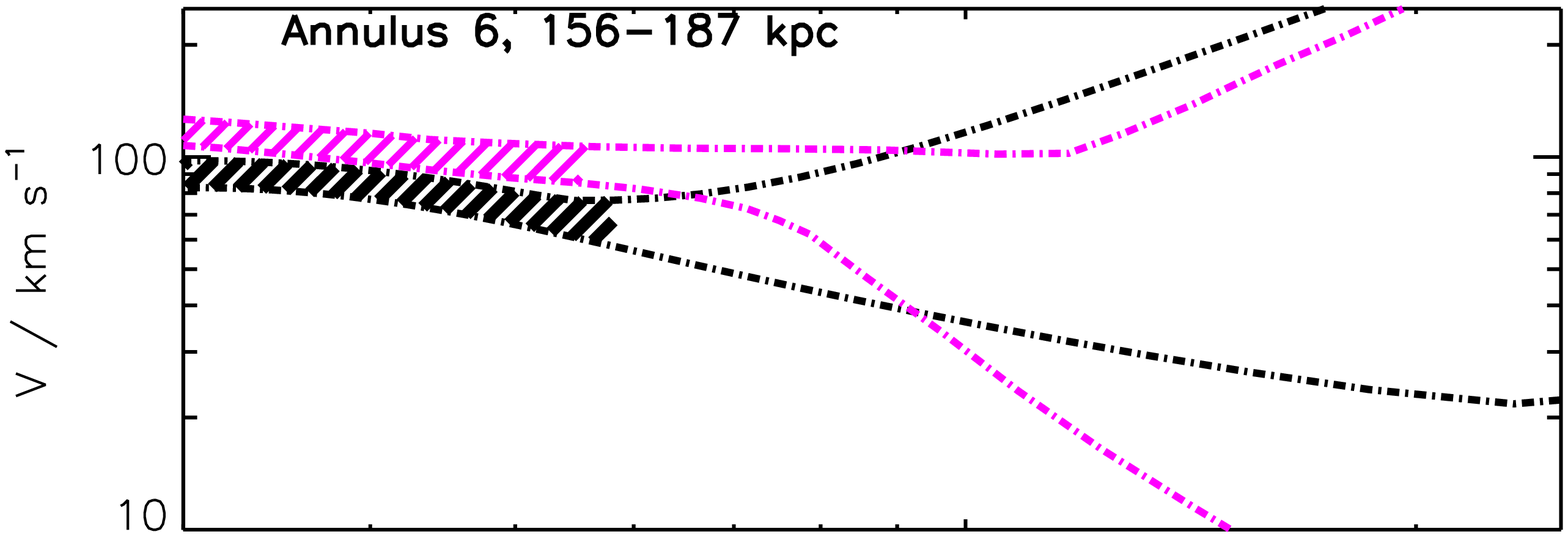, width=0.46\linewidth}
    \epsfig{file=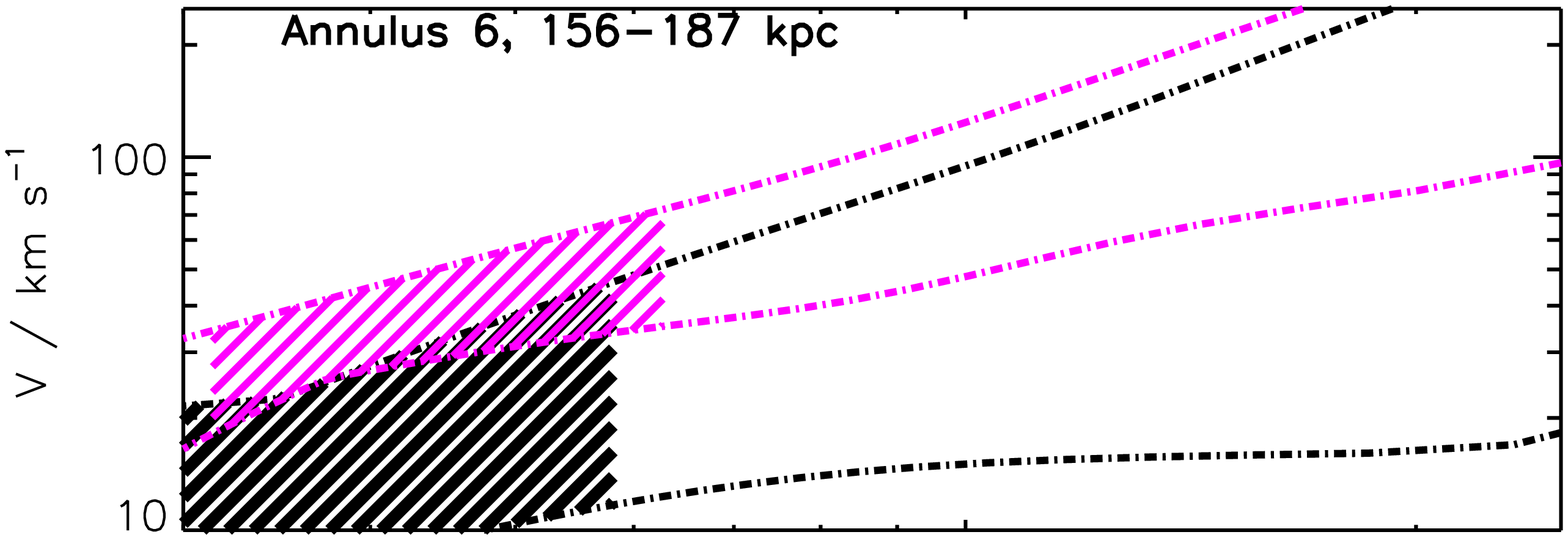, width=0.46\linewidth}
}
\vspace{-1.75cm}

\hbox{
\epsfig{file=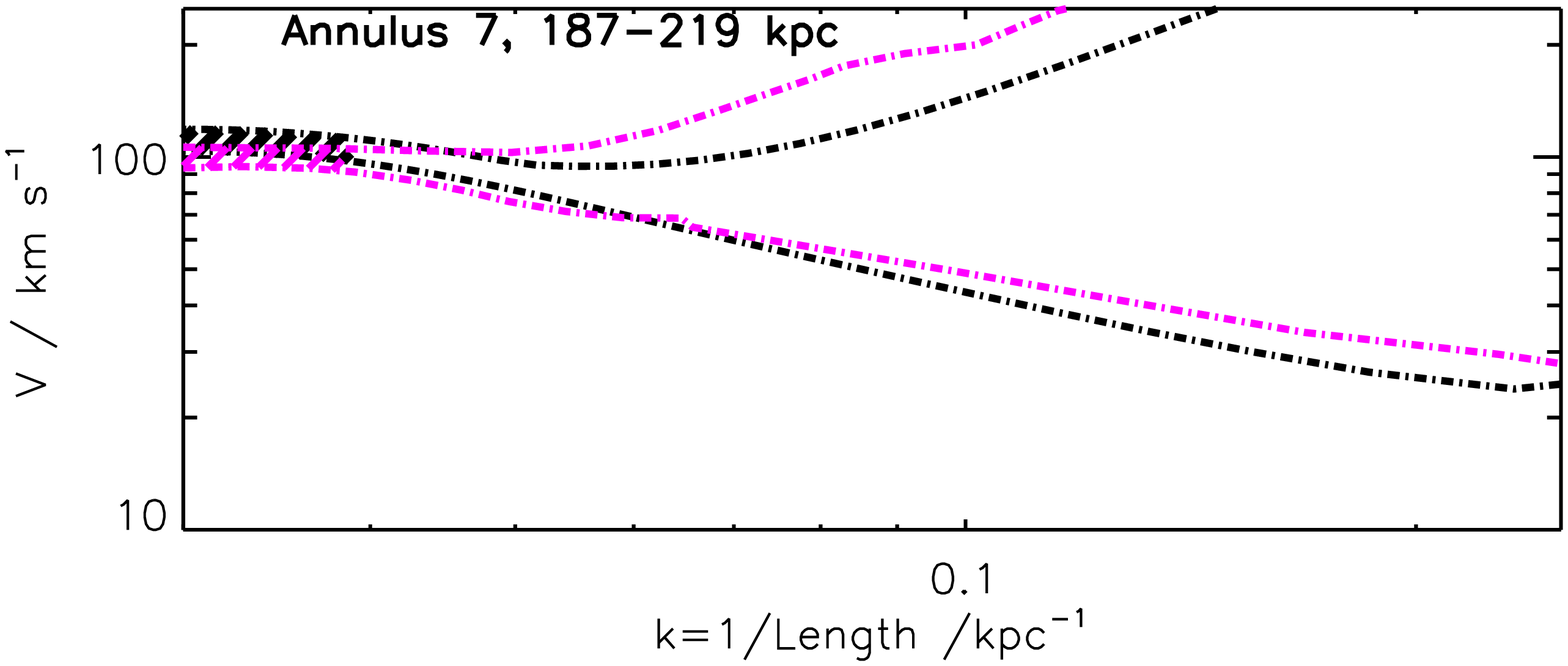, width=0.46\linewidth}
    \epsfig{file=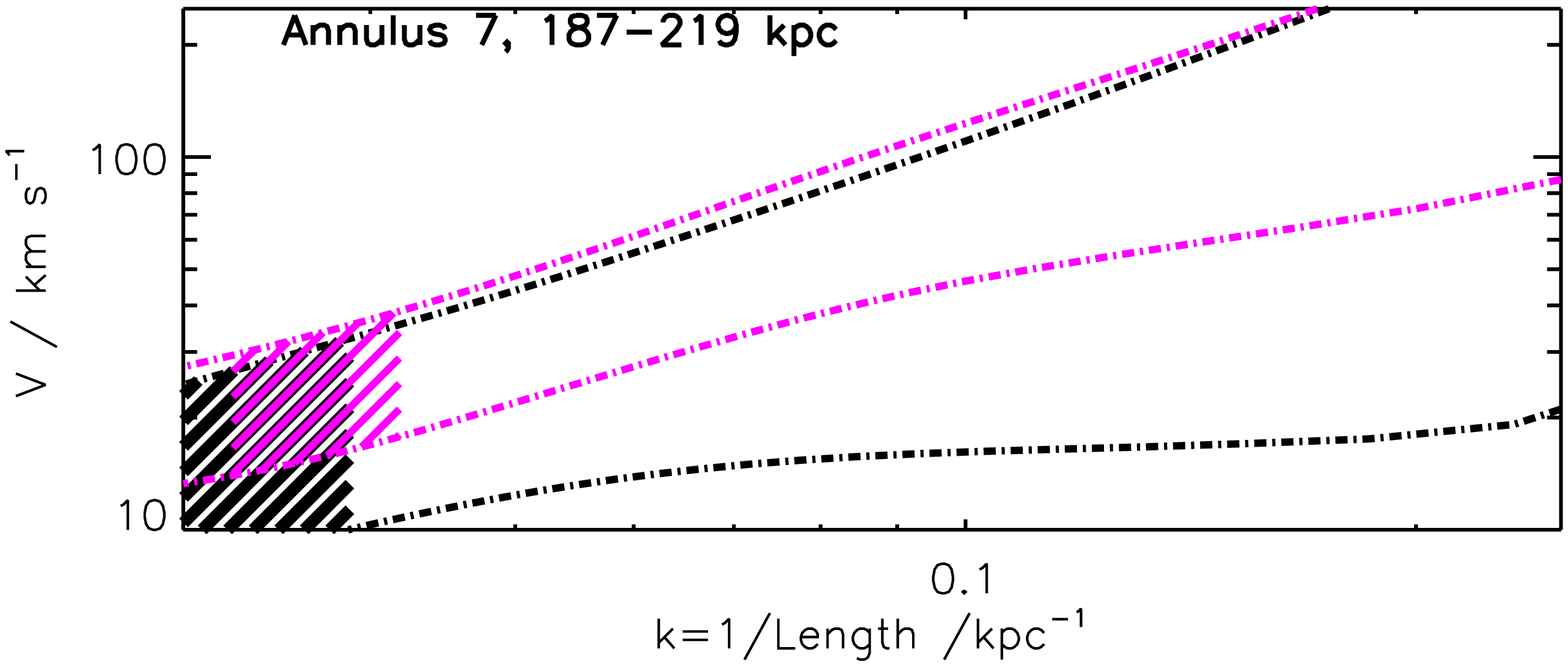, width=0.46\linewidth}
}

\vspace{-0.25cm}

\caption{Constraints on the velocity using the surface brightness fluctuation method of \citet{Zhuravleva2014}. The \emph{left} hand column shows the results when dividing the data 
by a simple beta model. The \emph{right} column shows the results when dividing by a patched beta model with $\sigma$=20 arcsec to remove the larger scale fluctuations. The dashed lines
show the upper and lower bounds for the constraints.
Magenta shows the results for the real Perseus data (the same as \citet{Zhuravleva2015}). Black shows the results of running this 
method on the fluctuations in the 1Ms simulated sloshing image. The shaded regions show the areas where the best constraints are made.  }
      \label{Velocities1}
  \end{center}
\end{figure*}

\begin{eqnarray}
P_{2D}(k)\approx 4P_{3D}(k)\int{|W(k_z)|^2dk_z},
\label{eq:klarge}
\end{eqnarray}
where $|W(k_z)|^2$ is the 1D power spectrum of the normalised emissivity, 
$\eta(z)$, along the line of sight

\begin{eqnarray}
W(k_z)=\int \eta(z) e^{-i2\pi z k_z} dz.
\label{eq:w}
\end{eqnarray}
where $\eta(z)$ is related to the density distribution $n^2_{0}(x,y,z)$ by

\begin{eqnarray}
\eta(x,y,z)=\frac{n^2_{0}(x,y,z)}{\int
  n^2_{0}(x,y,z')dz'}=\frac{n^2_{0}(x,y,z)}{I_0(x,y)}.
\end{eqnarray} 
For the small area of the cluster we consider the $x$ and $y$ dependence
can be neglected, yielding 
$\eta(x,y,z) \approx \eta(z)$.

\begin{figure*}
  \begin{center}
\hbox{
\epsfig{file=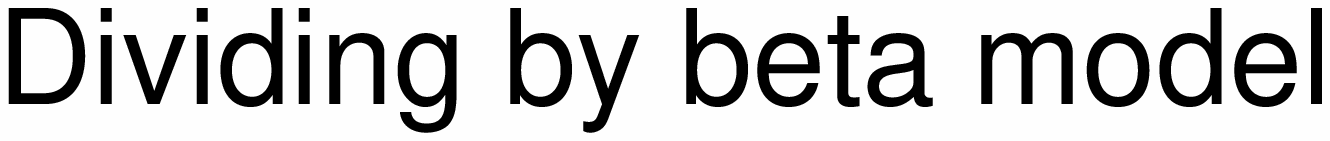, width=\linewidth}
}
\vspace{-0.7cm}

\hbox{
    \epsfig{file=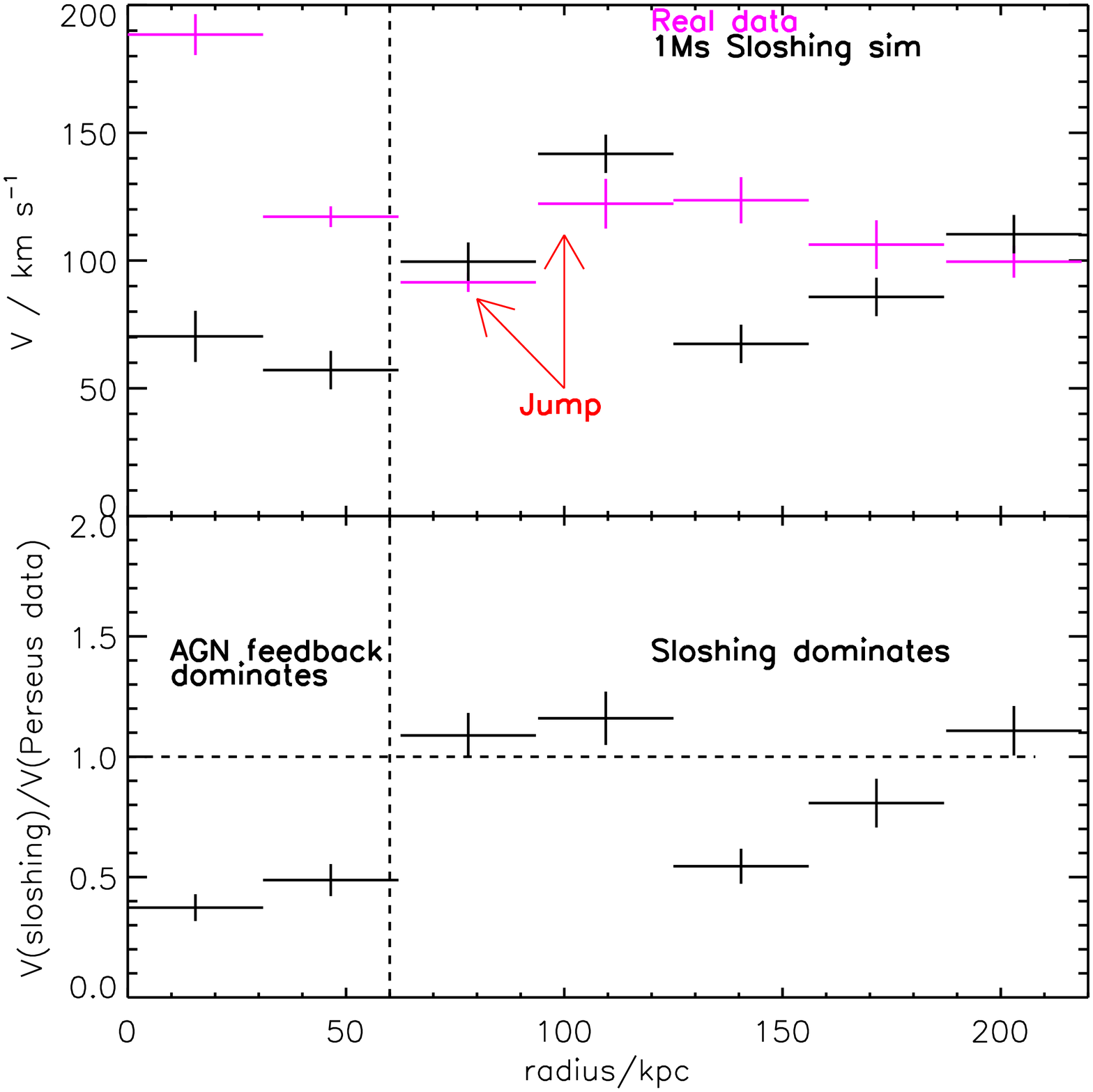, width=0.5\linewidth}
    \epsfig{file=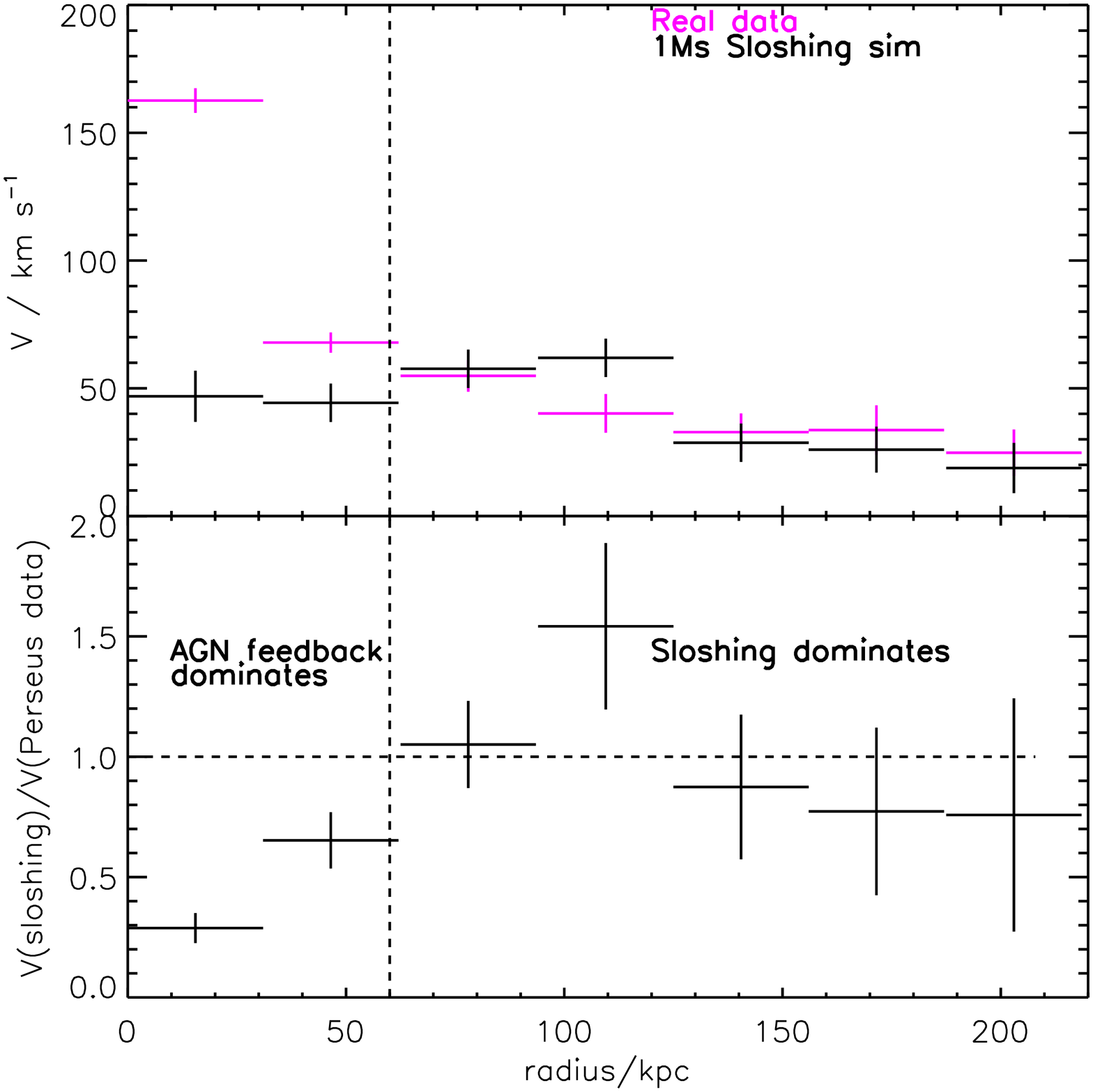, width=0.5\linewidth}

}
\caption{\emph{Top}:Comparing the velocity constraints on the length scale where the best constraints are made (i.e. the 
length scales of the shaded regions in Fig. \ref{Velocities1}) between the real Perseus data (pink) and the sloshing simulation (black) as a function of distance from the cluster core.
 \emph{Bottom}: The ratio of the sloshing velocities to the observed velocities in Perseus. Outside 60kpc these are in broad agreement. The \emph{left} hand figure is for the results obtained
when dividing the cluster by a beta model. The \emph{right} hand figure is obtained when the data are divided by a patched beta model with $\sigma$=20 arcsec.   }
      \label{Velocities_fraction}
  \end{center}
\end{figure*}

Following the approach of \citet{Zhuravleva2014} and 
\citet{Zhuravleva2015} we convert this 3D power spectrum into the one-component 
velocity, $V_{1,k}$ power spectrum. We expect the one-component 
velocity to be proportional to the amplitude of density 
fluctuations (for the stratified atmospheres of relaxed 
galaxy clusters, and within the inertial range of scales), $A_{3D}= \frac{\delta \rho_{k}}{\rho_{0}}$ at each length scale 
$l=1/k$, which leads to,

\begin{eqnarray}
A_{3D} = \frac{\delta \rho_{k}}{\rho_{0}} = \eta_{1} \frac{V_{1,k}}{c_s}
\end{eqnarray} 
where $c_s$ is the sound speed. It was found in \citet{Zhuravleva2014L} using cosmological 
simulations of galaxy clusters that the average value of the proportionality 
constant $\eta_{1}$ is consistent with unity ($\eta_{1} = 1 \pm 0.3$). 

The 3D power spectrum is related to the amplitude of the density fluctuations using

\begin{eqnarray}
A_{3D} = \sqrt{P_{3D}(k) 4 \pi k^{3}}
\end{eqnarray} 
allowing us to find the amplitude of the one-component velocity of the gas 
motions as a function of wavenumber ($k=1/l$) from

\begin{eqnarray}
V_{1,k} \approx c_s \sqrt{P_{3D}(k) 4 \pi k^{3}} / \eta_{1}
\end{eqnarray} 

These constraints on the one-component velocity in Perseus are shown for each annulus in the left hand column of Fig. 
\ref{Velocities1} as the pink profiles. We use the same correction for the effect of Chandra's PSF as that described in \citet{Zhuravleva2015}, 
and use the same thermodynamic profiles to obtain the same radial profile of sound speed. For each annulus we produce 1000 Poisson realisations given the number of photons in each annulus, 
and use the resulting scatter to compute the error due to Poisson noise. Following \citet{Zhuravleva2015} we add a 3 percent systematic uncertainty to the scatter of the Poisson noise power spectra to take 
into account small numerical and instrumental effects. Our results for the Perseus 
data are completely consistent with those 
reported in \citet{Zhuravleva2014,Zhuravleva2015}.

\subsection{Sloshing simulation analysis}
\label{sec:analysis_beta:sloshingsim}

\citet{Walker2017} found that the sloshing simulations of \citet{ZuHone2011,ZuHone2018} (in particular the time slice at 2.3Gyr from the onset of sloshing, for an initial uniform
thermal to magnetic pressure ratio of $\beta=200$) 
provide a good description of the cold front
swirl in the Perseus cluster, reproducing the size and distribution of the cold front as well as the location of a possible `bay' shaped by a
Kelvin-Helmholtz instability. These sloshing simulations also predict subtle surface brightness fluctuations on scales of 10s of kpc, as shown in 
the bottom right panel of Fig. \ref{Images}, where the contrast of the surface brightness fluctuations in the simulation have been enhanced by using 
a Gaussian Gradient magnitude (GGM) filter (\citealt{Sanders2016b}, \citealt{Walker2016}). 

We then folded the sloshing simulation through the mosaiced exposure map for the Chandra data (shown in the bottom left panel of Fig. \ref{Images}), 
to produce a mock Chandra observation with the same statistics as the actual Perseus data (top right panel of Fig. \ref{Images}). A constant background level was also added to the image, 
the same as that observed in the real Perseus data. The image was divided by the best fitting beta model to the azimuthally averaged surface brightness distribution, to find the residuals. 

The mock Chandra mosaic of the sloshing simulation was then divided into the same 7 annuli as the real data, from which the delta variance 
method was used to obtain the 2D power spectrum of the fluctuations. Following the same methods as for the real Perseus data described in section \ref{sec:analysis_beta:Perseus}, these 2D power spectra 
were converted into one-component velocities, and the resulting constraints are shown for each annulus as the black curves in the left hand column of Fig. \ref{Velocities1}. 

%We stress that these velocities infered from the sloshing simulation are not real. We are merely using this as a way to compare the amplitude of the power spectra to the results of \citet{Zhuravleva2014,Zhuravleva2015}.

\begin{figure*}
    \hbox{
    \epsfig{file=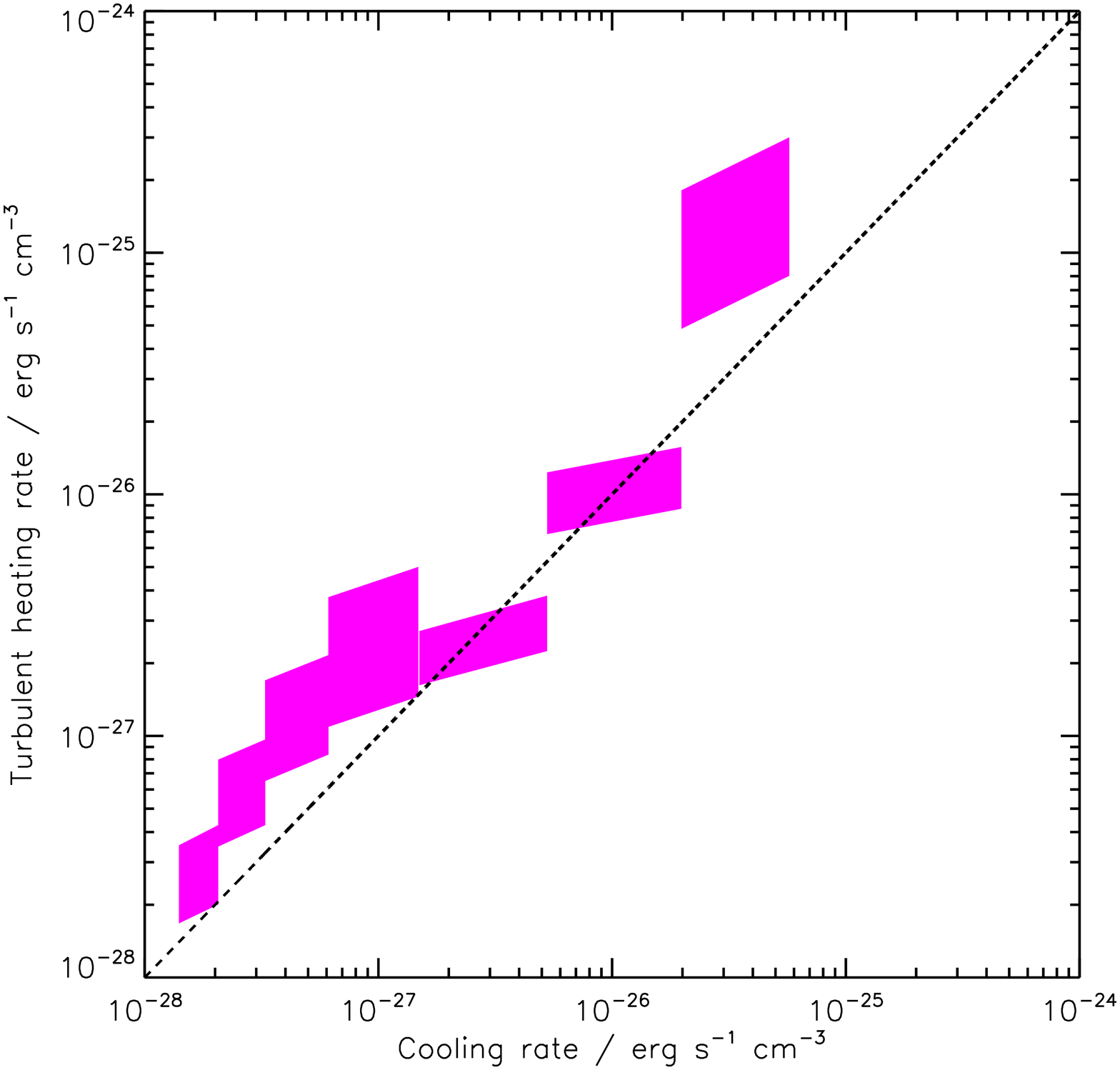, width=0.5\linewidth}
    \epsfig{file=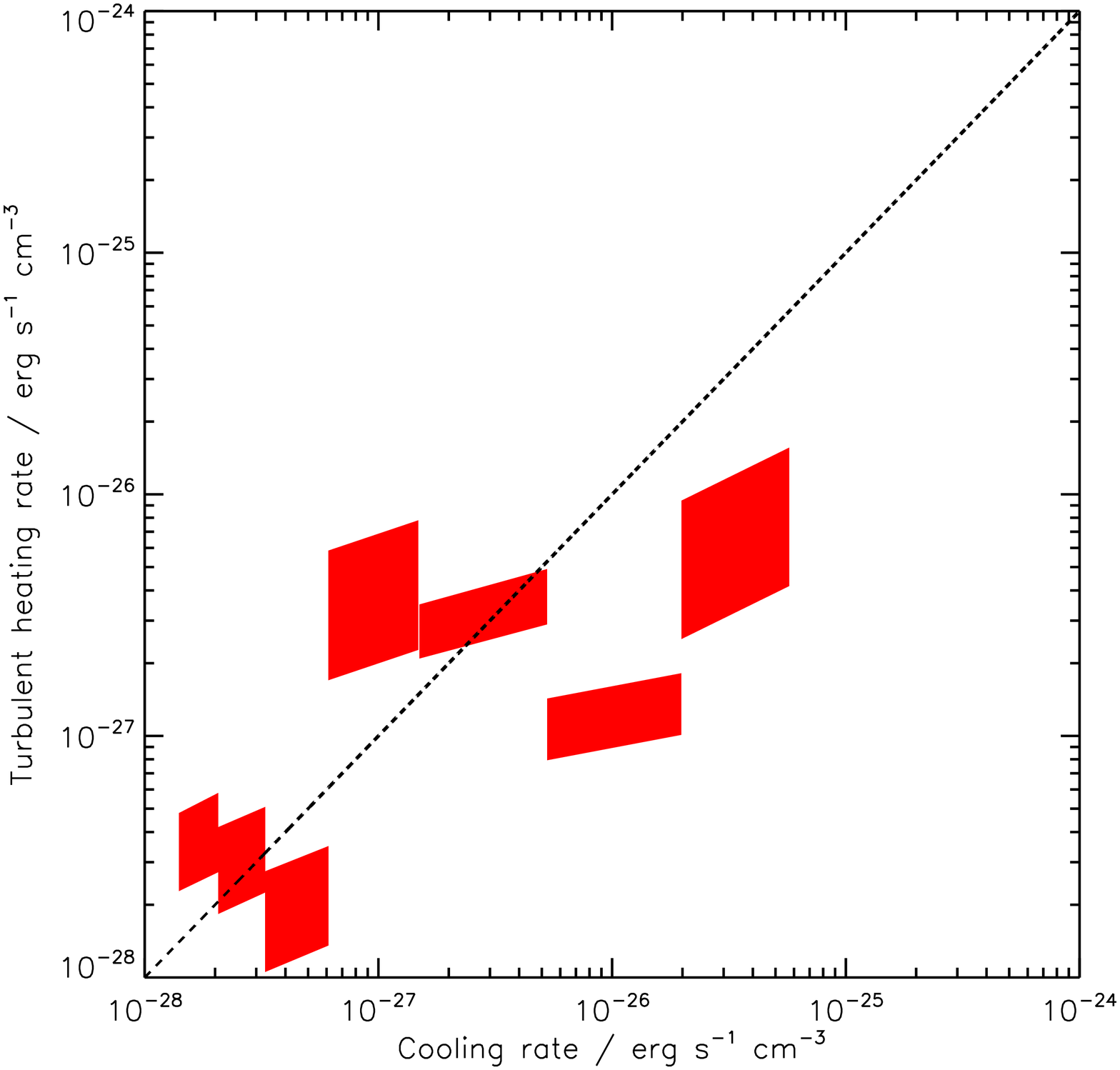, width=0.5\linewidth}

}
\caption{\emph{Left}:Turbulent heating versus cooling rate inferred by \citet{Zhuravleva2014} for the Perseus data
 \emph{Right}:The same plot but this time showing what happens if you just use the density fluctuations from the sloshing simulations. For the outer five annuli (the five
left most regions) the regions are either consistent with or above the cooling rate.    }
      \label{Heating_cooling_plot}
\end{figure*}

\subsection{Results from simple beta model removal}

From the velocity constraints shown in the left hand column of Fig. \ref{Velocities1}, we see that from the 3rd annulus outwards (i.e. outside 60kpc), the inferred velocity constraints (black curves) are
broadly consistent with those of the real Perseus data (pink curves). To show this more clearly, in the left panel of Fig. \ref{Velocities_fraction}, we show the profiles of the velocity constraints as a function of cluster radius, 
and then show the profile of the ratio between of the two. Outside 60kpc the velocity ratio is broadly consistent with unity, with the only exception of the 5th annulus. 

The profile for the real Perseus data (magenta points) shows a peculiar increase in the velocity from the third to the fourth annulus, marked by the red arrow, as found by 
\citet{Zhuravleva2015}. We see that 
this unusual behaviour would be readily explained by gas sloshing playing a dominant role in the observed power spectrum, as the `sloshing only' profile (black points) shows a similar
peak in the fourth annulus. This is caused by the main cold front edge being located in this fourth annulus. 

\citet{Zhuravleva2014} converted the observed one component-velocities into turbulent heating rates using  
\begin{eqnarray}
\Gamma_{diss} \sim 0.4 \rho v^{3} / l
\label{heating_rate_turbulence}
\end{eqnarray}

and then compared to the cooling rate for each annulus. As shown in the left hand panel of Fig. \ref{Heating_cooling_plot}
the inferred turbulent heating was found to match the cooling rate, or be slightly above it, in all the annuli studied. In this plot the cooling rate increases towards the core, so that moving from the left of the plot to the right we count down from annulus 7 (with the lowest cooling rate)
 to annulus 1 (with the highest cooling rate). To test the robustness of this, we repeated the same analysis on the sloshing fluctuations, 
converting the inferred one-component velocities into heating rates. The resulting profile is shown in the right hand panel of Fig. \ref{Heating_cooling_plot} with the red regions. 
Outside 60kpc, in the outer five annuli (the left most five regions), the inferred heating rate matches or is within 1-$\sigma$ of the cooling rate. This shows that, in principle, if the sloshing fluctuations are mis-interpreted as being due to 
AGN feedback, it is possible to incorrectly conclude that there is parity between heating rate from AGN feedback and cooling rates in the 60-220kpc region.  

\begin{figure*}
  \begin{center}
\hbox{
\hspace{1.0cm}
\epsfig{file=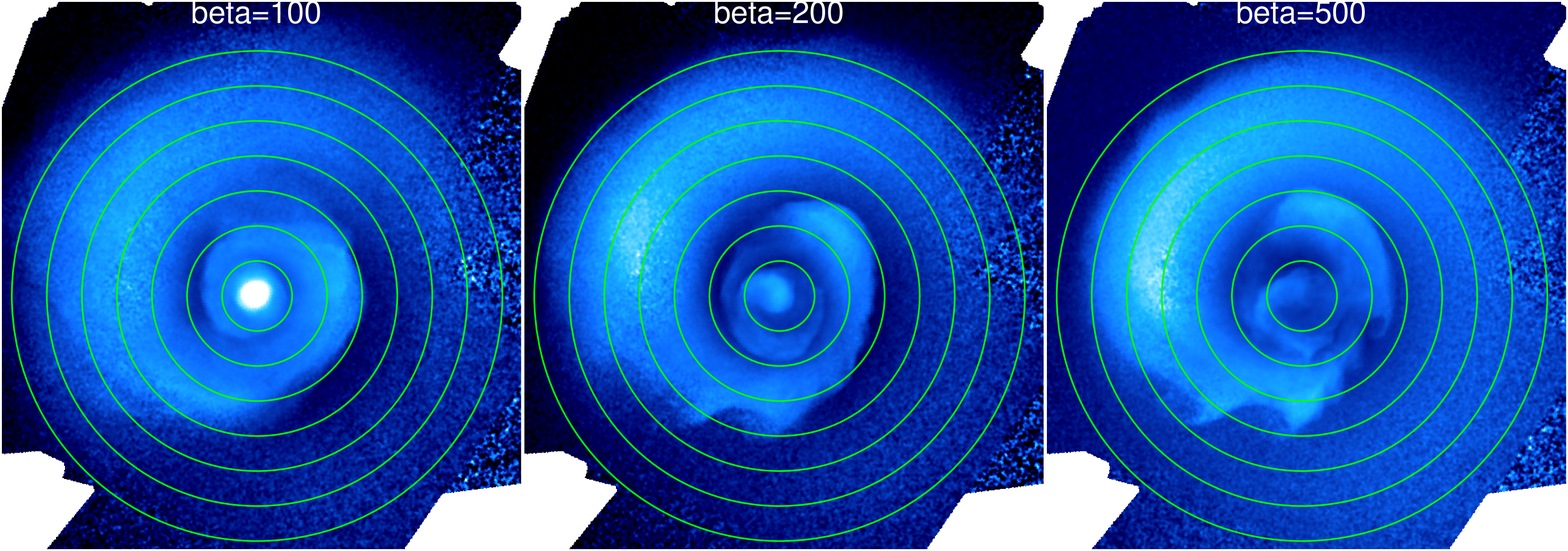, width=0.9\linewidth}
}
\vspace{-0.3cm}

\hbox{
    \epsfig{file=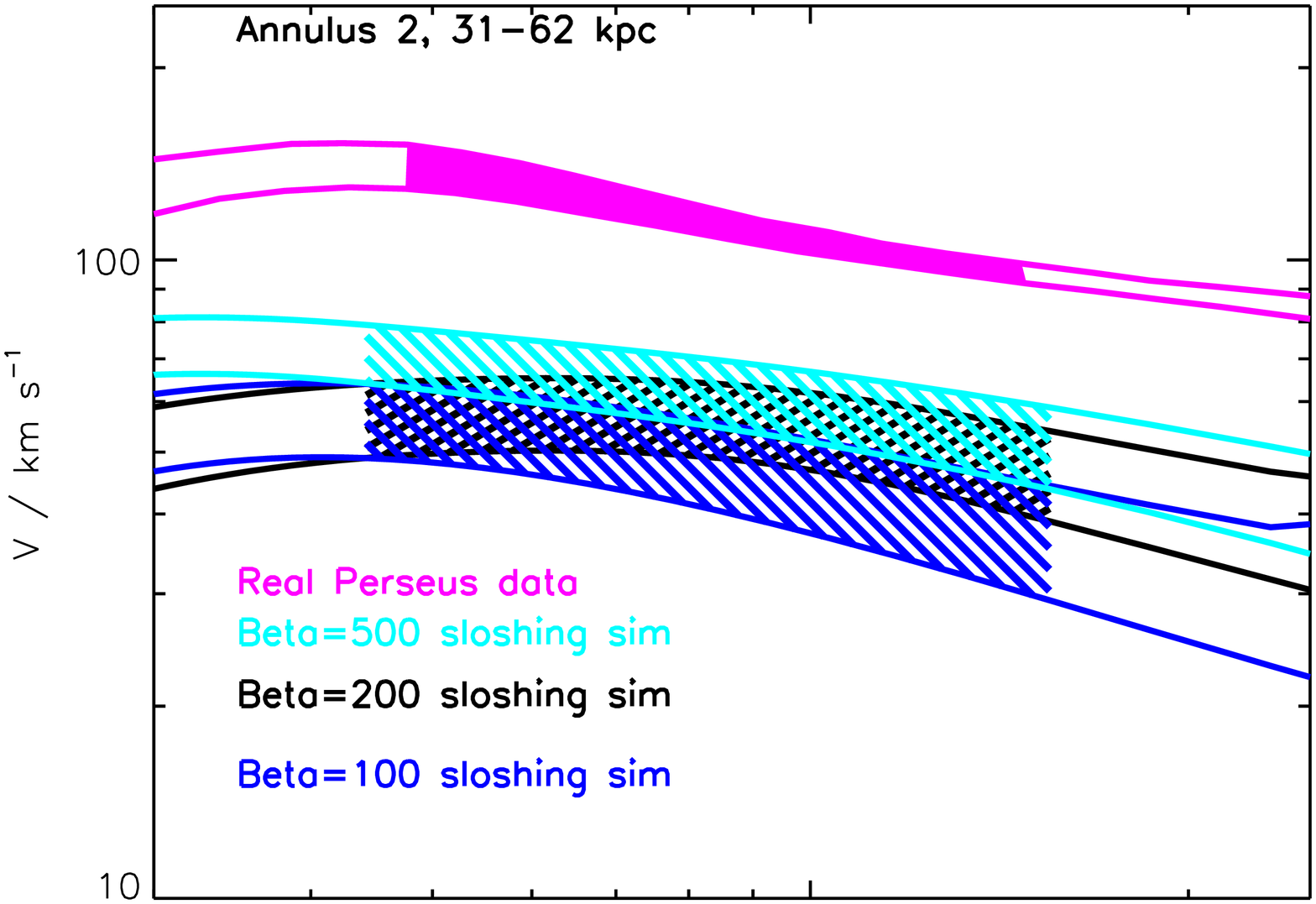, width=0.46\linewidth}
    \epsfig{file=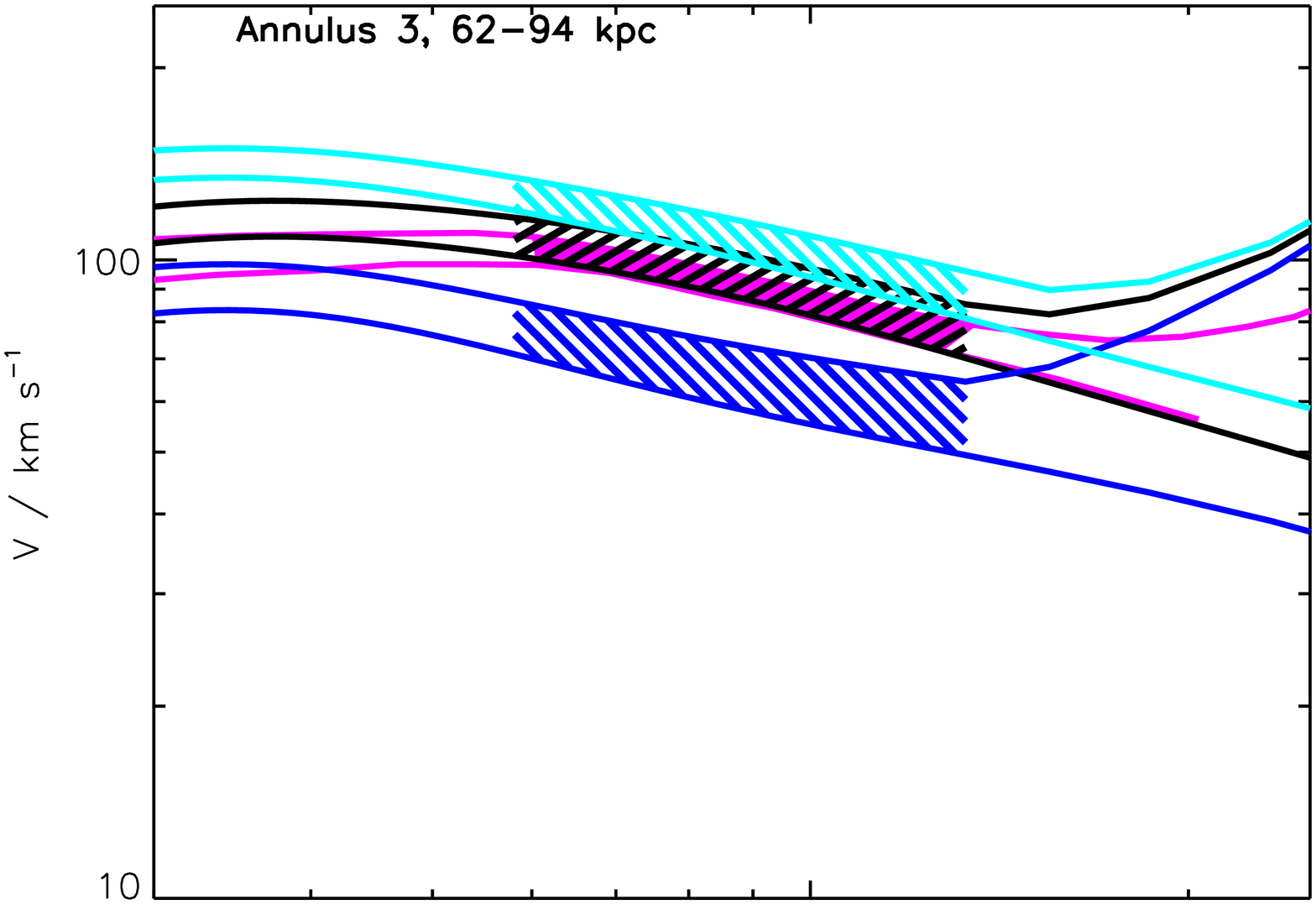, width=0.46\linewidth}
}
\vspace{-1.75cm}
\hbox{
    \epsfig{file=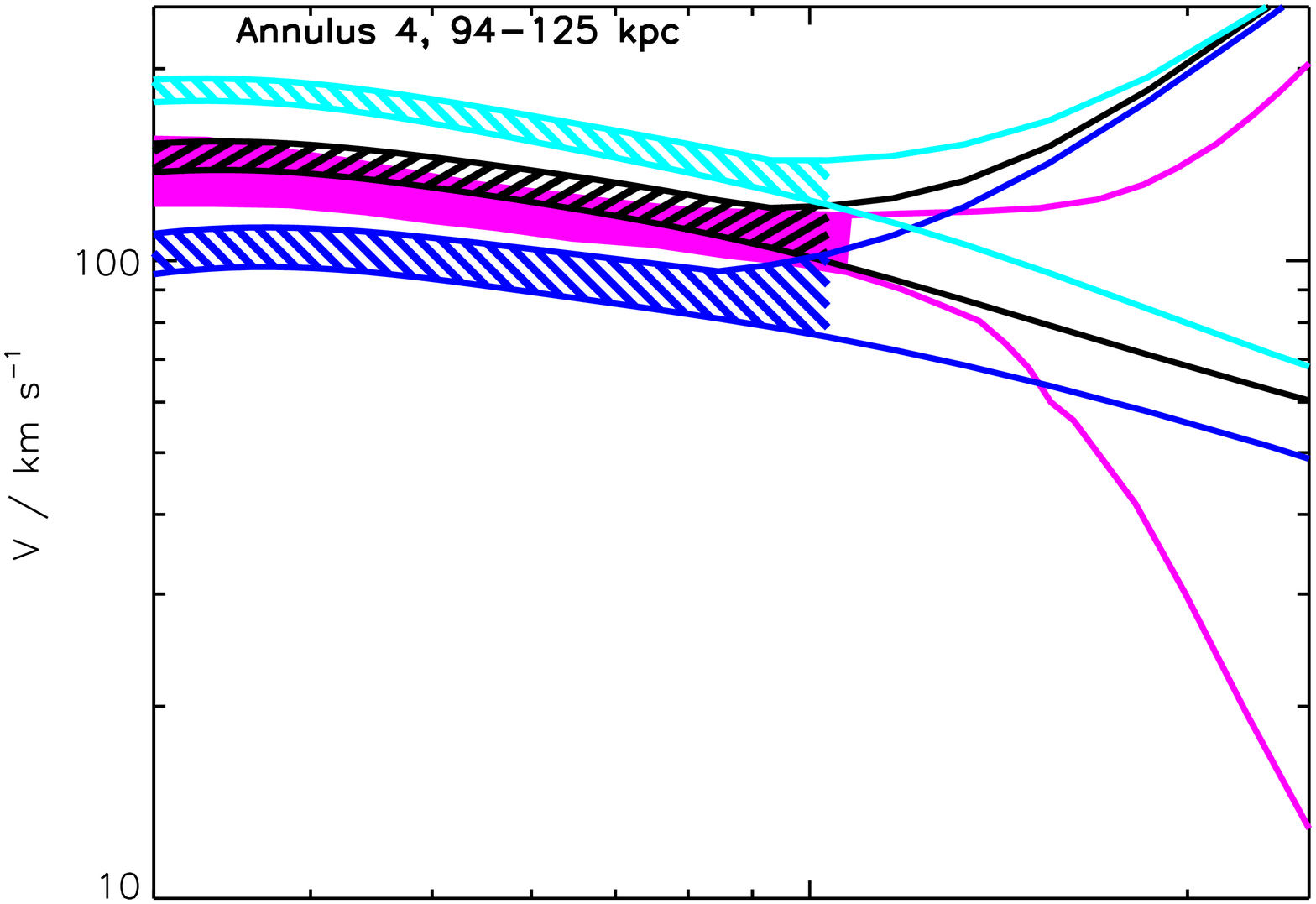, width=0.46\linewidth}
    \epsfig{file=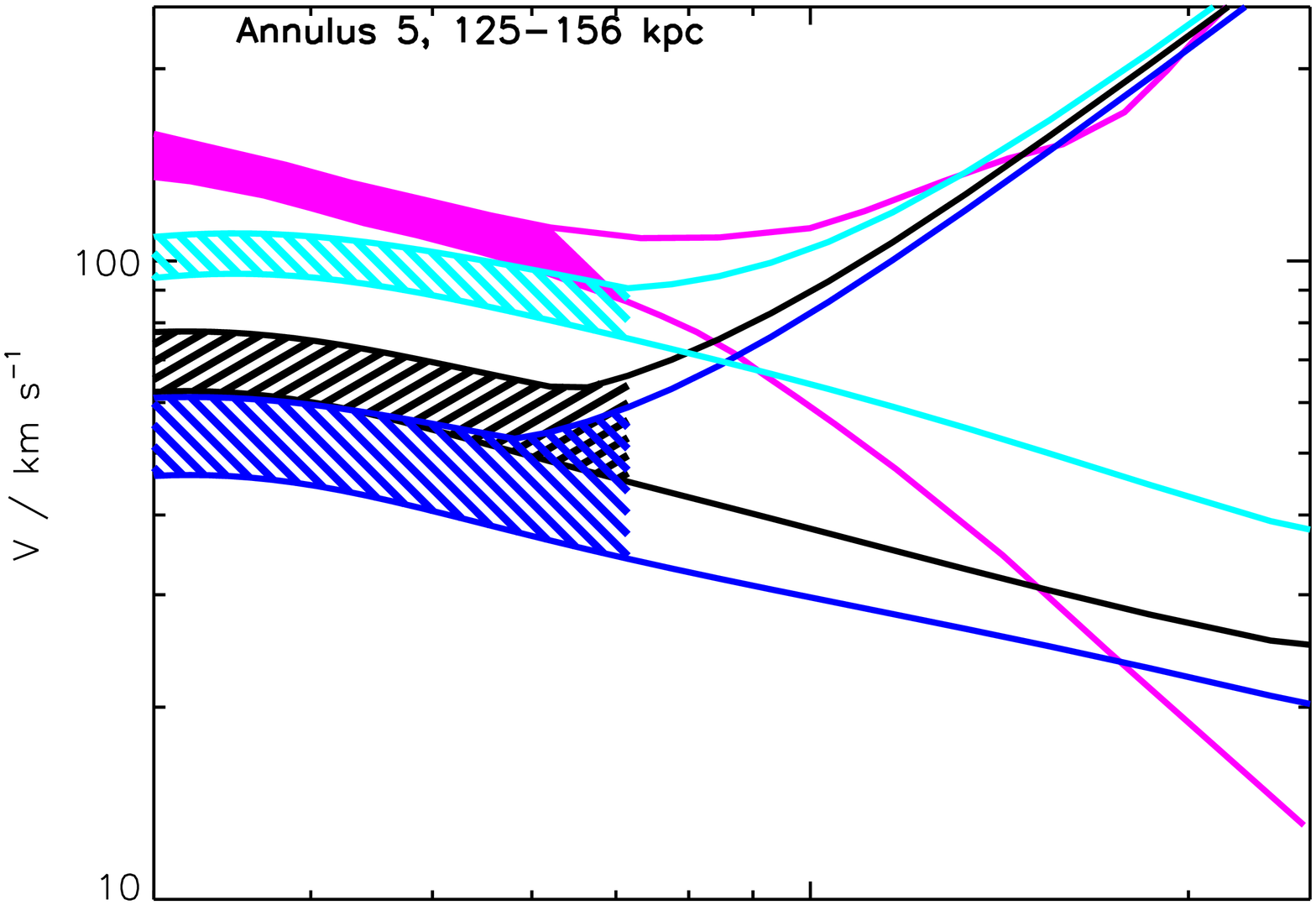, width=0.46\linewidth}
}
\vspace{-1.75cm}
\hbox{
    \epsfig{file=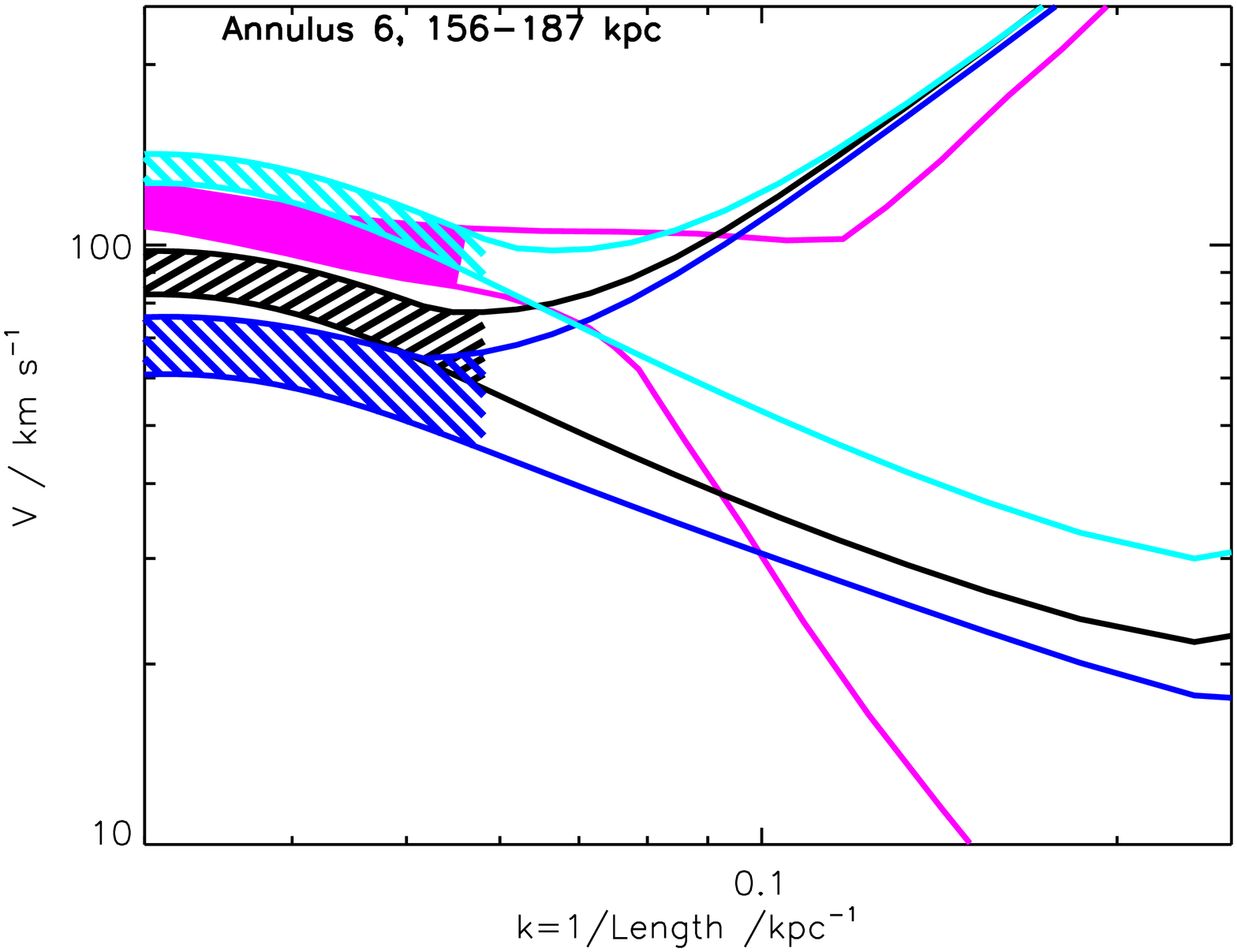, width=0.46\linewidth}
    \epsfig{file=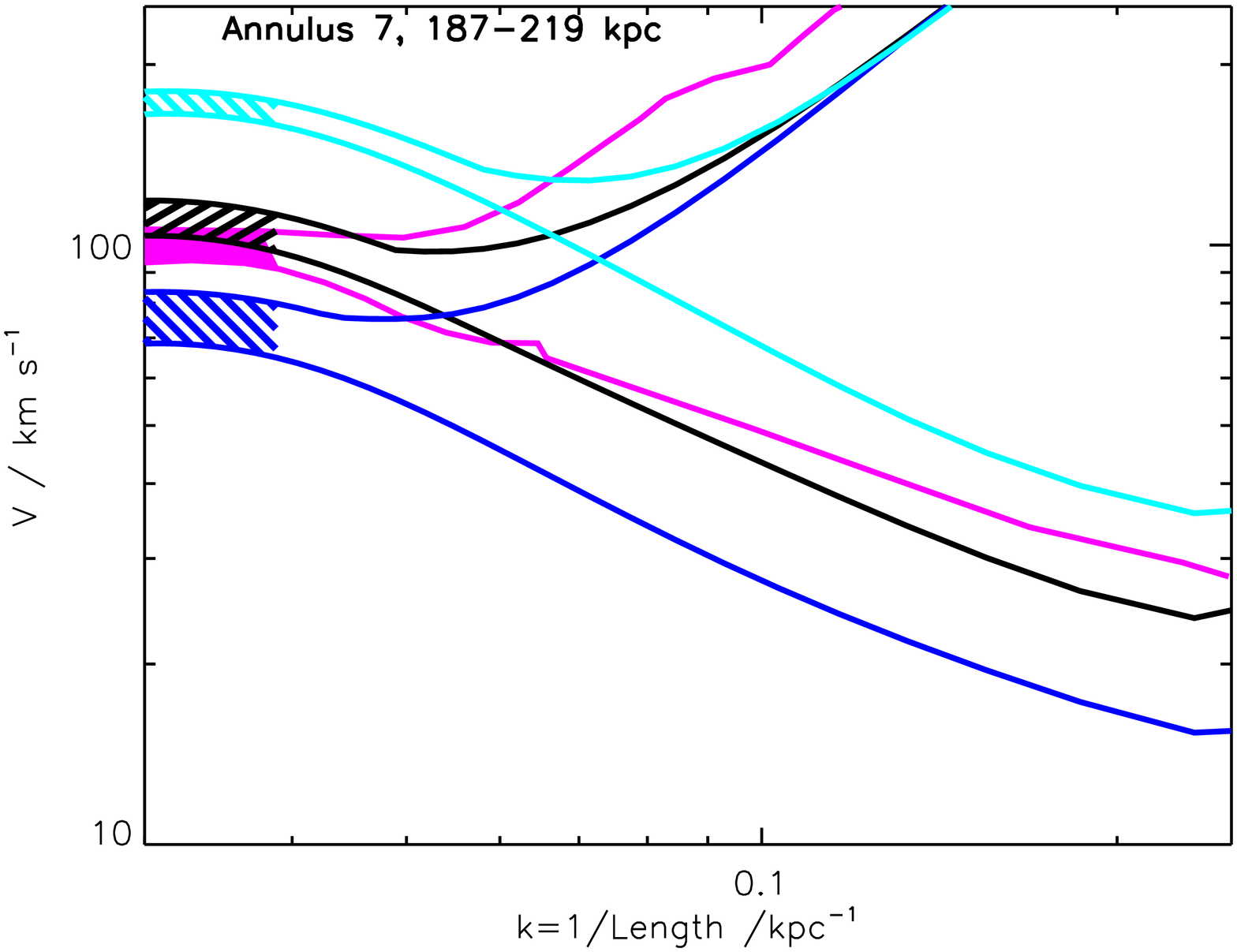, width=0.46\linewidth}
}

\caption{Comparing the constraints obtained from the sloshing simulations when different initial $\beta$s (the ratio of the thermal to magnetic pressure) are used in the sloshing
simulations. The $\beta$=200 simulations (black) provide the best match to the real data (pink). In all annuli, the simulations with a higher magnetic field ($\beta$=100, blue) have lower amplitude fluctuations,
while the simulations with a lower magnetic field ($\beta$=500, cyan) have higher amplitude fluctuations.   }
      \label{Different_betas}
  \end{center}
\end{figure*}

\section{Analysis: patched beta model removal method}
\label{sec:analysis_patchedbeta}

To assess the reliability of the small scale surface brightness fluctuations we measure, we repeat the analysis but this time use a `patched' beta model which accounts for structure on spatial scales of $\sigma=$20 arcsec. \citet{Zhuravleva2015} used 
this same method to show that, when a finely patched beta model is used, the larger scale fluctuations are indeed suppressed, but the smaller scale fluctuations ($0.1<k<0.2$) remain for case of the 
third annulus. This shows that these small scale fluctuations are real, and are not an artefact of the removal of the global cluster surface brightness distribution.

The results are shown in the right hand column of Fig. \ref{Velocities1}. The image at the top of the right hand column of Fig. \ref{Velocities1} shows the result of dividing the real Perseus image and the sloshing
simulation by a beta model patched on scales of $\sigma=$20 arcsec, showing that the larger scale surface brightness fluctuations are clearly suppressed. Moving down this column we then compare the constraints on the
surface brightness fluctuations in each annulus. We see again that, from the 3rd annulus outwards (i.e. from 60 kpc outwards) the level of the fluctuations seen in the real Perseus image is broadly consistent with the 
level seen in the sloshing simulations. Even the 5th annulus, which shows some discrepancy in the simple beta model removal method, is completely
consistent with the sloshing simulation. Due to the rapid decrease in surface brightness, from the 5th annulus outwards, the error bars become very large.

The radial profiles of the velocity constraints using the patched beta model are shown in the right hand panel of Fig. \ref{Velocities_fraction}, showing again that outside 60kpc the results for the real Perseus data 
are consistent with the sloshing simulation.

\section{Comparing sloshing simulations with different magnetic field strengths}

In the sloshing simulations, the initial magnetic field strength before sloshing begins has a significant impact on the development of structure. In \citet{Walker2017}, we compared the shape of the Perseus cold front and
the concave `bay' it features. We found that this shape could be most easily explained by a sloshing simulation with an initial uniform plasma $\beta$ (the ratio of the thermal pressure, $p_{th}$ to the magnetic pressure, $p_{B}$) of 200, as this leads to the 
development of a single large Kelvin-Helmholtz roll resembling the concave bay. When the initial magnetic pressure is higher ($\beta$=100), the large Kelvin-Helmholtz instabilities are suppressed and do not form. When the
magnetic field is lower ($\beta$=500), far more Kelvin-Helmholtz instability structure forms than is observed. This is shown in the image at the top of Fig. \ref{Different_betas}, which shows the same time slice for the sloshing
simulations with different initial uniform $\beta$s of 100, 200 and 500. 

Here we go further than the simple cold front shape arguments of \citet{Walker2017}, and compare the power spectra of the fluctuations in these different simulations with the real Perseus data. We follow the same method of folding the
simulations through the Chandra response for the same exposure time as the real observations, and then dividing the 
simulations by the best fitting azimuthally averaged beta model, and using the delta variance technique. The results for the annuli are shown in the plots in Fig. \ref{Different_betas}, where we compare the real Perseus
constraints (pink) with the $\beta=100$ (cyan), $\beta=200$ (black) and $\beta=500$ (blue) sloshing simulations. We omit the central annulus as this is so obviously dominated by AGN feedback. 

For all of the annuli, the $\beta=100$ simulations have smaller fluctuations than the $\beta=200$ simulations, which in turn have smaller fluctuations than the $\beta=500$ simulations. This can be trivially understood as a
decreasing magnetic field being less able to suppress the level of density fluctuations in the ICM. For most of the annuli (with the only exception being the 5th annulus), the $\beta=200$ simulation provides the best match 
to the real Perseus data.

Since the sloshing simulations from \citet{ZuHone2011} we use here are designed to investigate the effect on the cold fronts
of the cluster magnetic field over a large span of time (several Gyr), they do omit the effect of radiative cooling, which may affect the small scale
density fluctuations in the simulations. Previous works (\citealt{ZuHone2010}, 
\citealt{ZuHone2013a}) have shown that when radiative cooling is included in these sloshing simulations, a cooling catastrophe quickly develops in the innermost regions of the core. 
 The development of such cooling catastrophes when radiative 
cooling is included means that such simulations cannot be run very long, preventing a detailed study of the cold front structure which takes
several Gyr to develop.

% 
% \section{Data}
% \label{sec:data}
% \subsection{X-ray data}

\section{Conclusions}

$\bullet$ Outside 60kpc the level of surface brightness fluctuations is consistent with being entirely caused by the sloshing motions 
(Fig. \ref{Velocities_fraction}) in all except the fifth annulus studied. This main result is found both when the data are divided by a simple beta model, or by a more complex `patched' beta model
which removes structure on larger spatial scales. When a patched beta model is used, the power spectra in all of the annuli outside 60kpc match the sloshing simulation. \\
$\bullet$ If the level of density fluctuations 
is broadly consistent with sloshing outside 60kpc, AGN-generated turbulence is likely 
to be insufficient in combating cooling outside 60kpc. \\
$\bullet$ If the surface brightness fluctuations just from the sloshing simulation are converted into a turbulent heating rate, for all the annuli studied outside 60kpc (except the 5th annulus), the inferred turbulent 
heating rate is either consistent with or above the cooling rate (Fig. \ref{Heating_cooling_plot}, right hand panel). It is therefore 
possible to derive turbulent heating rates consistent with the cooling rates when the density fluctuations are 
entirely due to sloshing (and have nothing to do with AGN feedback). \\
% $\bullet$ When we subtract the simulated contribution from sloshing to the power spectrum in the 30-60kpc annulus (the radial range
% explored by Hitomi), the velocity is lowered by around 15 percent, moving it further away from the Hitomi constraint (top right panel 
% of Fig. \ref{Velocities}, blue region and curves).  \\
$\bullet$ The increase in the observed velocity in the Perseus data found by \citet{Zhuravleva2015} from the 60-90kpc to the 90-120kpc annuli
is readily explained by the profile shape for the sloshing cold fronts (see arrows on Fig. \ref{Velocities_fraction}). \\
$\bullet$ When we compare sloshing simulations with different initial plasma $\beta$s (the ratio of the thermal to magnetic pressure), we find that the simulation with 
an initial $\beta$ of 200 provides the best overall match to the real data. Simulations with lower initial magnetic fields ($\beta$=500) produce fluctuations that are generally larger than is observed,
while simulations with higher initial magnetic fields ($\beta$=100) produce fluctuations that are too small. This preferred $\beta$ agrees with that from a simple comparison of the shapes of the cold front structures 
from \citet{Walker2017}.\\
$\bullet$ Future, spatially resolved studies of Perseus with the high spectral resolution microcalorimeters on XARM/XRISM (\citealt{Ishisaki2018}) 
and Athena (\citealt{Barret2016}) will provide direct measurements of gas motions throughout the entire Perseus cluster core, further 
allowing the relative contributions from AGN feedback and gas sloshing to be disentangled.\\

\section*{Acknowledgements}
We thank the referee for helpful suggestions which
improved the paper. SAW was supported by an appointment to the NASA Postdoctoral Program at the
Goddard Space Flight Center, administered by the Universities Space Research
Association through a contract with NASA. ACF acknowledges ERC Advanced Grant 340442.
We thank John ZuHone for making the simulations shown in this paper publicly available. This
work is based on observations obtained with the \emph{Chandra} observatory, a 
NASA mission.
\bibliographystyle{mn2e}
\bibliography{Perseus_fluctuations}

% \appendix
% %
% \section[]{Observations}
% \label{appendix_obs}
% 
% 
% 
% 
% 
% \begin{table*}
% \begin{center}
% \caption{Chandra data used in this paper.}
% \label{obsdata}
% \leavevmode
% \begin{tabular}{l| l|l | l | l | l} \hline \hline
% Object & Obs ID&Exposure (ks)&RA&Dec&Start Date\\ \hline
% Perseus & 3209&95.77&03 19 47.60&+41 30 37.00&2002-08-08\\
% & 4289&95.41&03 19 47.60&+41 30 37.00&2002-08-10\\
% & 4946&23.66&03 19 48.20&+41 30 42.20&2004-10-06\\
% & 4947&29.79&03 19 48.20&+41 30 42.20&2004-10-11\\
% & 6139&56.43&03 19 48.20&+41 30 42.20&2004-10-04\\
% & 6145&85.00&03 19 48.20&+41 30 42.20&2004-10-19\\
% & 4948&118.61&03 19 48.20&+41 30 42.20&2004-10-09\\ 
% & 4949&29.38&03 19 48.20&+41 30 42.20&2004-10-12\\
% & 6146&47.13&03 19 48.20&+41 30 42.20&2004-10-20\\
% & 4950&96.92&03 19 48.20&+41 30 42.20&2004-10-12\\
% & 4951&96.12&03 19 48.20&+41 30 42.20&2004-10-17\\
% & 4952&164.24&03 19 48.20&+41 30 42.20&2004-10-14\\
% & 4953&30.08&03 19 48.20&+41 30 42.20&2004-10-18\\ \hline
% 
% \end{tabular}
% \end{center}
% \end{table*}

\end{document}